%% file: paper.tex
\documentclass[fleqn,usenatbib,useAMS]{mnras}
\usepackage{amssymb,amsmath,calrsfs}
\usepackage{graphicx,epstopdf}
\usepackage{soul,xcolor}

\newcommand{\beq}{\begin{equation}}
\newcommand{\enq}{\end{equation}}
\newcommand{\m}[1]{\boldsymbol{#1}}
\newcommand{\bra}[1]{\left#1}
\newcommand{\ket}[1]{\vphantom{\sqrt{0}}\right#1}

\title[Long-term evolution of the magnetosphere]{Long-term evolution of the force-free twisted magnetosphere of a magnetar}
\author[T. Akg\"{u}n et al.]
{T.~Akg\"{u}n$^1$\thanks{E-mail: akgun@astro.cornell.edu}, P.~Cerd\'{a}--Dur\'{a}n$^2$, J.A.~Miralles$^1$, and J.A.~Pons$^1$ 
\\$^1$Departament de F\'{i}sica Aplicada, Universitat d'Alacant, Ap. Correus 99, 03080 Alacant, Spain
\\$^2$Departament d'Astronomia i Astrof\'{i}sica, Universitat de Val\`{e}ncia, Dr. Moliner 50, 46100, Burjassot, Val\`{e}ncia, Spain}

\begin{document}
\label{firstpage}
\pagerange{\pageref{firstpage}--\pageref{lastpage}}
\maketitle

\begin{abstract}
	We study the long-term quasi-steady evolution of the force-free magnetosphere of a magnetar coupled to its internal magnetic field. We find that magnetospheric currents can be maintained on long timescales of the order of thousands of years. Meanwhile, the energy, helicity and twist stored in the magnetosphere all gradually increase over the course of this evolution, until a critical point is reached, beyond which a force-free magnetosphere cannot be constructed. At this point, some large-scale magnetospheric rearrangement, possibly resulting in an outburst or a flare, must occur, releasing a large fraction of the stored energy, helicity and twist. After that, the quasi-steady evolution should continue in a similar manner from the new initial conditions. The timescale for reaching this critical point depends on the overall magnetic field strength and on the relative fraction of the toroidal field. The energy stored in the force-free magnetosphere is found to be up to $\sim 30\%$ larger than the corresponding vacuum energy. This implies that for a $10^{14}$\,G field at the pole, the energy budget available for fast magnetospheric events is of the order of a few $10^{44}$\,erg. The spindown rate is estimated to increase by up to $\sim 60\%$, since the dipole content in the magnetosphere is enhanced by the currents present there. A rough estimate of the braking index $n$ reveals that it is systematically $n < 3$ for the most part of the evolution, consistent with actual measurements for pulsars and early estimates for several magnetars.
\end{abstract}

\begin{keywords}
	magnetic fields -- MHD -- stars: magnetars -- stars: magnetic field -- stars: neutron.
\end{keywords}

\input{introduction}

\input{results}

\section*{Acknowledgements}
This work is supported in part by the Spanish MINECO grants AYA2015-66899-C2-1-P, AYA2015-66899-C2-2-P, the grant of Generalitat Valenciana PROMETEOII-2014-069 and by the New Compstar COST action MP1304.

\bibliographystyle{mnras}
\bibliography{references}

\label{lastpage}
\end{document}

%% file: introduction.tex

\section{Introduction}
The presence of a global twist (a toroidal component) in the magnetospheres of strongly magnetized neutron stars (NSs) is inferred from observations of quiescent magnetar spectra \citep{2008ApJ...686.1245R,2015SSRv..191..315M,2017arXiv170300068K}, which show features that are attributed to resonant cyclotron scattering due to the presence of magnetospheric currents. Such a twist could possibly be maintained by helicity transfer from the stellar interior, implying that the magnetosphere of a non-rotating star is not necessarily current-free (though, still force-free). Further observational evidence for the existence of twisted magnetospheres is provided by magnetar outbursts, and in particular by the outburst decay behavior. The dissipation of large magnetospheric twists on timescales of years has been invoked to explain the magnetar outburst decay properties \citep{2000ApJ...543..340T,2002ApJ...574..332T,2007ApJ...657..967B,2009ApJ...703.1044B,2013ApJ...777..114B}. In this paper, we aim to understand how, and at which rate, helicity is transferred from the NS crust to the exterior, and what are the conditions for the magnetosphere to remain stable on long timescales or to undergo a global reorganization (flare).

In \cite{2016MNRAS.462.1894A}, hereafter referred to as \emph{Paper I}, we presented axisymmetric, force-free models for the magnetosphere of a magnetar. In this paper, we continue our investigation, adapting these models as a boundary condition for the long-term internal evolution of the magnetic field given by the code described in \cite{2012CoPhC.183.2042V}. In Paper I, we found that the force-free magnetosphere can store up to about 25\% more energy with respect to the vacuum model. We also found that the presence of currents in the magnetosphere can lead to an overestimation in the value of the surface dipole moment by a comparable amount. This excess also defines the available energy budget in the event of a fast, global magnetospheric reorganization in the magnetic field structure, commonly associated with magnetar flares. As also noted in Paper I, there is a maximum twist ($\varphi_{\rm max}$) allowed in the magnetosphere. We were unable to find solutions when the maximum twist reaches a certain critical value ($\varphi_{\rm max}\sim 1.5$\,rad) after exploring a wide range of parameters. Similar results were obtained by \cite{2002ApJ...574..332T} using an analytical self-similar solution ($\varphi_{\rm max} =\pi$) and by \cite{2017MNRAS.468.2011K} in force-free configurations ($\varphi_{\rm max} \sim 1.6$ to $3$). Resistive MHD simulations of the dynamics of twisted magnetospheres were performed by \cite{1994ApJ...430..898M}, applied to the disruption of coronal arcades, reaching similar conclusions ($\varphi_{\rm max} \sim 3.2$) and by \cite{2012ApJ...754L..12P,2013ApJ...774...92P}, in the context of magnetar magnetospheres ($\varphi_{\rm max}\sim 3$, see footnote\footnote{Different authors use different definitions for the twist angle. In \cite{2002ApJ...574..332T}, \cite{2012ApJ...754L..12P,2013ApJ...774...92P} and \cite{2016MNRAS.462.1894A}, it is the north--south twist, while in \cite{1994ApJ...430..898M} and \cite{2017MNRAS.468.2011K}, it is the north--equator twist. In this work, we use the north--south twist and multiply by $2$ when referring to  the results of authors using a different convention.}). Increasing the twist further would result in a sudden disruption of the magnetospheric loops. We also note that magnetars, the main target for this study, have relatively slow rotation, therefore we can safely neglect its effects. In that case, the pulsar equation reduces to the Grad--Shafranov equation, which has been widely studied both in astrophysics and in plasma physics, but has only limited analytic solutions available.

Other recent numerical solutions for NS magnetospheres are given in \cite{2014MNRAS.437....2G,2014MNRAS.445.2777F,2015MNRAS.447.2821P}; and \cite{2017MNRAS.468.2011K}. \cite{2014MNRAS.437....2G} and \cite{2015MNRAS.447.2821P} have presented numerical solutions by solving the Grad--Shafranov equation in the interior and the exterior continuously. Similarly, \cite{2014MNRAS.445.2777F} impose barotropic equilibrium in the core, Hall equilibrium in the crust and force-free equilibrium outside. However, the crustal field evolution (due to the Hall and Ohmic terms) takes the field away from such simple equilibria (either barotropic or Hall),  within the characteristic timescales of interest in the long-term evolution. Stable stratification (due to composition gradients) and the elastic response of the crust can balance small deviations and can help to maintain some quasi-equilibrium, which, however, is no longer given through the strict requirement imposed by the Grad--Shafranov equation.

The dynamical evolution of twisted magnetospheres has been studied, for example, in \cite{2012ApJ...754L..12P,2013ApJ...774...92P} and \cite{2017ApJ...844..133C}. In these models, shearing of the crust (which is added by hand and limited to a certain region on the surface) is seen to lead to a series of (slow) magnetospheric expansion and (sudden and energetic) reconnection events, consistent with observations of flares and bursts. As the twist increases, the energy stored in the magnetosphere grows, and the field lines tend to inflate \citep[as also noted by][]{1995ApJ...443..810W}. Expansion of the magnetosphere also leads to the opening up of more field lines beyond the light cylinder. Therefore, they conclude that increasing the magnetospheric twist should also strongly affect (increase) the spindown rate, and could explain irregularities already detected in some objects (for example SGR 1806$-$20, SGR 1900$+$14 and XTE J1810$-$197). In particular, a twist larger than $\sim 1$\,rad is expected to cause significant changes in the spindown \citep{2009ApJ...703.1044B,2012ApJ...754L..12P}. However, such large twists are also expected to be unstable \citep[and references therein]{2002ApJ...574.1011U}, and should result in the ejection of a fraction of the energy in the form of a plasmoid.

In our case, the interior magnetic field is determined by the long-term evolution due to the Hall and Ohm terms in the crust. The evolution in the core would involve additional ambipolar diffusion terms, complicating the picture due to their highly non-linear nature, and is not considered here. We also assume that the magnetosphere \emph{instantaneously} reaches a static equilibrium solution. In other words, we do not solve for the detailed dynamics of the magnetosphere, which would happen on very fast (Alfv\'{e}n) timescales (much shorter than the long-term evolution timescales due to the Hall and Ohm effects in the crust). At each step in the evolution, the magnetosphere is assumed to quickly dissipate any transient perturbations, and adjust nearly instantaneously to a new force-free solution, coupling to the magnetic field at the surface. In all previous works on magneto-thermal evolution \citep[e.g.][]{2012CoPhC.183.2042V,2013MNRAS.434..123V} the boundary condition imposed on the magnetic field was that of a vacuum (current-free) magnetic field. The internal evolution gives the radial component of the magnetic field at the surface, and the surface boundary condition returns the tangential component of the field compatible with a vacuum solution. In this paper, with the new implementation, we are able to generalize this external boundary condition, which now allows for the presence of currents (and twist) in the exterior. The vacuum case is retrieved as a special case for zero toroidal field.

This paper is structured as follows: in \S\ref{section_theory} we present a short overview of the theoretical and technical details of the model and its implementation; in \S\ref{section_results} we present results for the magnetospheric evolution driven by the internal magnetic field evolution; and in \S\ref{section_conclusions} we discuss the potential implications of our findings.

\section{Theory}\label{section_theory}
We assume that the equilibrium structure of the magnetosphere is determined entirely by the magnetic field, and that the pressure and inertia of the plasma are negligible there. We thus model the magnetosphere in terms of a force-free, but not necessarily current-free, magnetic field. We consider the case of magnetars where rotation is relatively slow (with typical periods in the range of 2 to 12 seconds) and can be safely neglected (as the corresponding light cylinder has a radius of over $10^5$\,km, well beyond the region of interest of a few stellar radii, i.e.\ $\lesssim 100$\,km).

Throughout this paper, we will employ the same notation as in Paper I. (More in-depth discussion of force-free fields can be found there, while here we only present a minimal overview of the subject.) An axisymmetric magnetic field can be written in terms of two stream functions as, in spherical coordinates $(r,\theta,\phi)$,
	\beq
	\m{B} = \m\nabla P \times \m\nabla\phi + T \m\nabla\phi \ ,
	\enq
where $P(r,\theta)$ is the poloidal stream function and $T(r,\theta)$ is the toroidal stream function. In a static axisymmetric fluid, the Lorentz force cannot have an azimuthal component, so $T$ must be a function of $P$. In a barotropic fluid, the force density must further be expressible as the gradient of a potential. This gives rise to the so-called Grad--Shafranov equation, which determines the magnetic field structure in a non-rotating plasma,
	\beq
	\triangle_{\rm GS} P + G(P) = \varrho \varpi^2 F(P) \ .
	\enq
Here, $G=T T'$, $\varrho$ is density and $\varpi$ is cylindrical radius. Force-free implies $F=0$, while current-free further requires $G=0$ (a more detailed discussion is given in Paper I). In our model, the magnetospheric toroidal field is confined in a region near the equator, while near the poles (where the field lines extend to the light cylinder) the field is current-free. At the stellar surface we impose continuity of the field lines, while at a specified external radius (usually taken to be around 10 stellar radii) we smoothly match to a current-free (vacuum) solution. The magnetosphere model is scalable, in the sense that its structure is independent of the overall amplitude of the magnetic field, and only depends on the relative ratio between the poloidal and toroidal components (for a given toroidal stream function profile).

We do not model the dynamics of the magnetosphere (which would happen on the timescale of seconds) but rather its equilibrium. We are interested in the long-term evolution (of the order of kyr to Myr) governed by the induction equation in the interior \citep{2012CoPhC.183.2042V,2013MNRAS.434..123V}. In this model, the magnetosphere will instantaneously adjust to the surface magnetic field. One of the key issues is determining the functional dependence $T(P)$ between the toroidal and poloidal functions in the magnetosphere. The internal evolution is, in general, inconsistent with such a relation, and even if the initial field satisfies such a condition, the subsequent evolution causes the poloidal and toroidal functions to evolve separately, without maintaining any specific functional relationship between the two. More specifically, even if one were to start with a Hall equilibrium, which is also given by a (different) Grad--Shafranov equation with the toroidal and poloidal stream functions having some functional dependence, the Ohm term would eventually take the system out of this equilibrium and break down this dependence. Thus, a crucial point in applying the boundary conditions at the surface is determining the functional relation between the two stream functions, and enforcing this relation there.

To properly understand the evolution of $T(P)$ at the surface, one must consider the dynamical reaction of the magnetosphere to perturbations and how it relaxes to a force-free solution. \cite{2014MNRAS.443.1416G} performed ideal MHD simulations of the propagation of internal torsional oscillations to the magnetosphere in magnetars. The initial unperturbed poloidal magnetic field lines were symmetric with respect to the equator and were perturbed by a toroidal component that was either symmetric or antisymmetric with respect to the equator. In the symmetric case, perturbations were able to twist the magnetosphere while in the antisymmetric case, the perturbations were reflected at the surface and the magnetosphere remained untwisted. From the point of view of the force-free condition, the Grad--Shafranov equation only admits solutions for symmetric perturbations. That is, for a single field line labeled by a given value of $P$, both footprints in the northern and southern hemispheres must have the same value of $T$. For antisymmetric perturbations, the northern and southern footprints have different values of $T$ (with opposite signs), which is incompatible with the force-free condition, unless $T=0$. This argument  remains valid in general, when there is no symmetry with respect to the equator, by considering symmetric or antisymmetric perturbations of a field line with respect to its footprints. An alternative way of understanding the effect is realizing that the two footprints of a magnetic field line are connected through the magnetosphere and they cannot evolve independently from each other. If one of the footprints moves, it creates a current that flows through the magnetosphere to the other footprint. If both footprints move, the net current in the magnetosphere will be the addition of the currents coming from both footprints. In the case of antisymmetric perturbations, these two currents are equal but with opposite signs and hence the total current cancels out. Therefore, the appropriate boundary conditions at the surface have to ensure that the function $T(P)$ is not multivalued, which effectively filters the antisymmetric component of the perturbations, but allows symmetric perturbations to twist the magnetosphere.

There are two cases in which the previous arguments do not hold. The first case is for open magnetic field lines connected to the light cylinder, which for magnetars are typically those within $\sim 0.01$\,rad from the axis. Currents can then flow freely and independently in the north and south poles. The second case is when the timescale in which the footprints move is shorter than the Alfv\'{e}n crossing time along the magnetic field line \citep{2014MNRAS.443.1416G}. In this case, perturbations are reflected and trapped inside the star \citep{2014MNRAS.441.2676L,2014MNRAS.443.1416G}. For magnetic field lines inside the light cylinder, the Alfv\'{e}n crossing time is of the order of $\sim 10$\,ms \citep{2014MNRAS.443.1416G}, which is many orders of magnitude shorter than all characteristic timescales in the slow magneto-thermal evolution.

\subsection{Internal evolution}
The evolution of the internal magnetic field in an NS crust is given through Faraday's law of induction, with a generalized Ohm's law for the electric field,
	\beq
	\begin{split}
	\partial_t \m{B} & = - c \m\nabla \times \m{E} \\
	& = - \m\nabla \times \bra{[} \eta \m\nabla \times \m{B}
	+ f_H (\m\nabla \times \m{B}) \times \m{B} \ket{]} \ .
	\end{split}
	\enq
Here, $\eta$ is the magnetic diffusivity, related to the electrical conductivity $\sigma$ by  $\eta = c^2/4\pi\sigma$, and we have defined the Hall coefficient $f_H = c/4\pi  e n_{\rm e}$, where $n_{\rm e}$ is the electron number density and $e$ the elementary charge. For brevity, here we omit further details, and we refer to \cite{2013MNRAS.434..123V} for a detailed description of the evolution equation, timescales and observational implications, and to \cite{2012CoPhC.183.2042V} for details on the numerical code for the internal evolution. Throughout this paper, we use an NS model with a mass of $M_\star = 1.4 M_\odot$ and a radius of $R_\star = 11.6$\,km.

\subsection{Initial magnetic field}\label{section_initial}
We start our simulation with an internal magnetic field of the form described in \cite{2013MNRAS.433.2445A}. The poloidal field is an analytic construction for a dipolar field in a non-barotropic star, i.e.\ it is not a solution of the Grad--Shafranov equation. It satisfies regularity and continuity conditions, and smoothly joins to a dipole vacuum field at the surface.

On the other hand, the toroidal field is of the same form as in Paper I, and is confined within the magnetic surface defined by the critical field line corresponding to $P = P_{\rm c}$,
	\beq
	T(P) \propto \left\{
	\begin{aligned}
	& (P-P_{\rm c})^\sigma & \mbox{for} \ P \geqslant P_{\rm c} \ , \\
	& 0 & \mbox{for} \ P < P_{\rm c} \ .
	\end{aligned}
	\right.
	\label{T_of_P}
	\enq
Regularity of the magnetospheric current requires $\sigma \geqslant 1$ (as discussed in Paper I). In this work, we always start with $\sigma=1$. As the field evolves, the function $T(P)$ must adapt, and terms of higher order will appear.

We choose an intermediate value for the initial $P_{\rm c}$ somewhere in the interval from $0$ (corresponding to the pole) up to the maximum value $P_{\rm o}$ of the poloidal function at the surface (initially at the equator). Both $P_{\rm c}$ and $P_{\rm o}$ will change throughout the evolution. The presence of a toroidal field in the exterior at the start of our simulation implies that the field structure of the magnetosphere is determined by solving the Grad--Shafranov equation right from the very first step, and clearly will not match continuously to the vacuum values initially assumed for the internal poloidal field at the stellar surface. More specifically, at the start of the simulation, the radial and azimuthal components of the magnetic field, $B_r$ and $B_\phi$, are continuous, while the meridional component $B_{\theta}$ is not. Therefore, initially there are surface currents, which are then rapidly redistributed and smoothed over (in a transient phase) within the first couple of decades of the evolution (first few tens of time steps), and the field lines are subsequently seen to make a smooth transition from the interior to the exterior.

The initial toroidal field has a maximum amplitude of $B_{\phi,\rm max}$ somewhere in the interior of the star, in the equatorial plane. Note that, this maximum does not correspond to the \emph{neutral line}, where the poloidal stream function has a maximum and the poloidal field vanishes. In other words, while the contours of the stream functions $P$ and $T$ coincide, and correspond to the poloidal field lines, the contours of $B_\phi$ are obviously different than these, and only coincide with the poloidal field lines for the contour $P=P_{\rm c}$ (or equivalently, $T=0$).

The initial \emph{surface} poloidal field has a maximum at the pole, of amplitude ${\rm max}[B_{\rm pol}(R_\star,\theta)] = B_r(R_\star,0) \equiv B_{\rm pole}$, while the absolute maximum of the poloidal field is located at the center of the star. We define the ratio of the toroidal field amplitude to the poloidal field amplitude at the start of the simulation (at $t=0$) as
	\beq
	\mathfrak{f} \equiv \frac{B_{\phi,\rm max}}{B_{\rm pole}} \ .
	\label{fraction}
	\enq
This ratio controls the relative strength of the toroidal field with respect to the poloidal field. In all of the simulations presented in this paper, we set the maximum poloidal field strength at the surface ($B_{\rm pole}$) to $10^{14}$\,G. The maximum toroidal amplitude is then given through the fraction $\mathfrak{f}$ defined by the above equation. In particular, we will refer to the case with the initial values $P_{\rm c} = 0.5 P_{\rm o}$ and $\mathfrak{f}=0.6$ as \emph{model A} for the rest of this paper.

\subsection{Boundary conditions}\label{section_boundary}
At each timestep, the radial component of the magnetic field at the surface is used to calculate the poloidal function $P(R_\star,\theta)$, while the toroidal function $T(R_\star,\theta)$ is derived from the azimuthal component evaluated at the last internal numerical cell. In the force-free magnetosphere, $T$ and $P$ should be functions of one another, so we must impose this condition. We carry out a non-linear best fit to specify an analytic form for $T(P)$. In order to avoid discontinuities in the function, we assume a relation of the form
	\beq
	T_{\rm fit}(P) = \left\{
	\begin{aligned}
	& a_1 (P - a_2) + a_3 (P - a_2)^2 & \mbox{for} \ P \geqslant a_2 \ , \\
	& 0 & \mbox{for} \ P < a_2 \ ,
	\end{aligned}
	\right.
	\label{Tfit}
	\enq
where $a_i$ are three free parameters to be determined. We restrict ourselves to linear and quadratic terms, although we have tried different functional forms and the results do not change qualitatively. The amplitudes of the two terms are controlled by the parameters $a_1$ and $a_3$, while $a_2$ is the critical value of the poloidal stream function ($P_{\rm c}$) marking the boundary between the force-free and vacuum regions.

Note that this fitting function for $T$ is not a monotonic function of $P$, and may have a maximum somewhere in the interval $P_{\rm c} < P < P_{\rm o}$, with $P_{\rm o}$ being the maximum value of the poloidal function at the surface. In that case, the derivative $T'(P)$ will go through zero and change sign. Since the current for a force-free field is proportional to the magnetic field through (see Paper I)
	\beq
	\frac{4\pi\m{J}}{c} = T'(P) \m{B} \ ,
	\label{current}
	\enq
this implies that the current will also continuously decrease to zero at some magnetic surface, and then reverse direction. In principle, this should not cause any problems for the equilibrium structure of the magnetosphere, though its implications for stability are less clear.

Using the prescription for $T_{\rm fit}(P)$, and the boundary conditions $P(R_\star,\theta)$, we construct a magnetosphere by numerically solving the Grad--Shafranov equation iteratively (as outlined in Paper I). We calculate the resulting meridional and toroidal components of the magnetic field ($B_\theta$ and $B_\phi$) immediately above the surface, consistently with the external magnetosphere. These two components are then used as external boundary conditions for the next time step in the interior. This more general boundary condition allows for currents to flow through the surface into the magnetosphere and back into the interior, allowing for the transfer of helicity, twist and energy into/from the magnetosphere.

\subsection{Numerical implementation}
In this section, we discuss some technical details of the implementation. Starting with an initial magnetic field, we apply the new boundary condition for the force-free twisted magnetosphere at each step of the internal magnetic field evolution. We first construct a best fit for the toroidal function at the surface by minimizing $\chi^2$, defined in the usual way,
	\beq
	\chi^2 = \sum_i^{N_\theta} \frac{\bra{[}T_i - T_{\rm fit}(P_i)\ket{]}^2}{N_\theta T_{\rm o}^2} \ ,
	\enq
where we ascribe equal weights to all of the $N_\theta$ points on the surface, and $T_{\rm o}$ is the amplitude of $T$. (When $P$ is measured in units of some $P_{\rm o}$, $T$ is given in units of $T_{\rm o} = P_{\rm o}/R_\star$.) The function $T_{\rm fit}$ is given by equation (\ref{Tfit}). Because of the non-linear dependence on the parameters, we employ the Levenberg--Marquardt method, which minimizes $\chi^2$ iteratively \citep[see, for example,][]{Press:1993:NRF:563041}. This functional form of $T(P)$ then holds throughout the magnetosphere. Following the transient period of a couple of decades where the system adjusts to the initial conditions (by dissipating initial surface currents), $\chi^2$ settles to a value typically below $10^{-3}$.

For the stellar interior, we typically use a $200 \times 200$ staggered grid, with some of the quantities only calculated in intermediary cell points, and then extrapolated to the normal grid, as described in \cite{2012CoPhC.183.2042V}. For the magnetosphere, we use an external radius of 10 stellar radii, so in order to maintain a comparable radial resolution to the interior, we need significantly more points at a correspondingly heavier computational cost. In practice, we find that a resolution of 800 radial points in the magnetosphere gives us satisfactory accuracy while still being fairly fast on a typical desktop computer.

Since the previous solution is used as an estimate in the new time step, the Grad--Shafranov solver often converges in just a couple of iterations. We evaluate the convergence by calculating the $\chi^2$ for the $k$th iteration (with respect to the previous one) over the two dimensional (2D) grid in the magnetosphere, as defined in equation (22) of Paper I, which we reproduce here as
	\beq
	\left\{\chi^2\right\}_k = \sum_{i,j}^{N_r,N_\theta}
	\frac{\left(\left\{P_{i,j}\right\}_k - \left\{P_{i,j}\right\}_{k-1}\right)^2}{N_r N_\theta P_{\rm o}^2} \ .
	\enq
$N_r$ and $N_\theta$ are the sizes of the radial and angular grids, respectively. Typically, this $\chi^2$ is well below $10^{-7}$, except near the point beyond which convergence is no longer achieved and no solutions are found. (For more details on the Grad--Shafranov solver, refer to Paper I.)

%% file: results.tex

	\begin{figure*}
		\centerline{\includegraphics[width=0.33\textwidth]{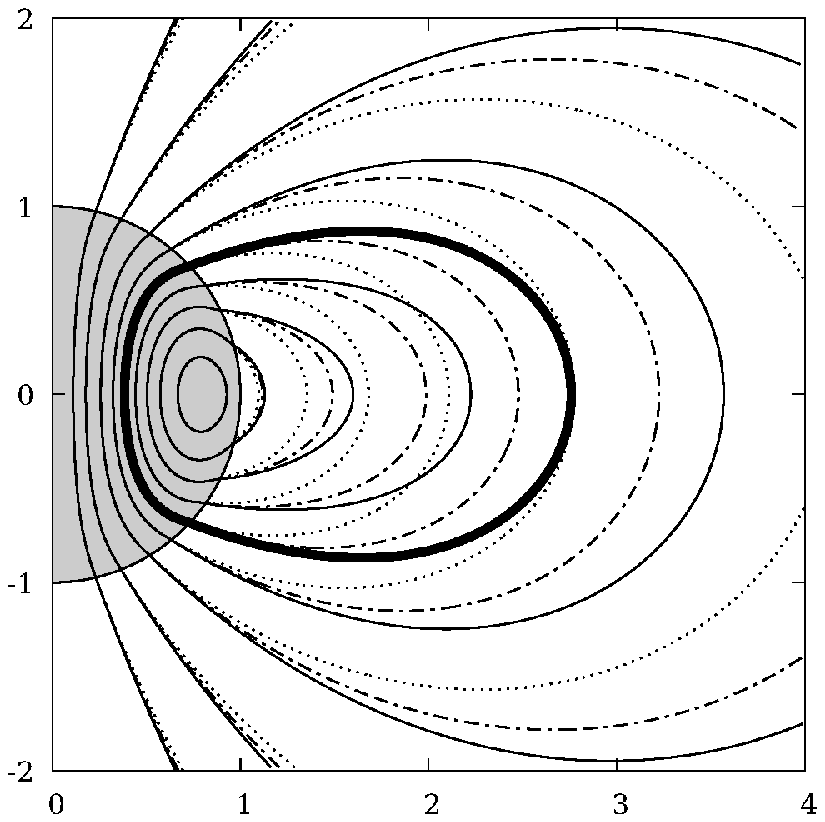}
			\includegraphics[width=0.33\textwidth]{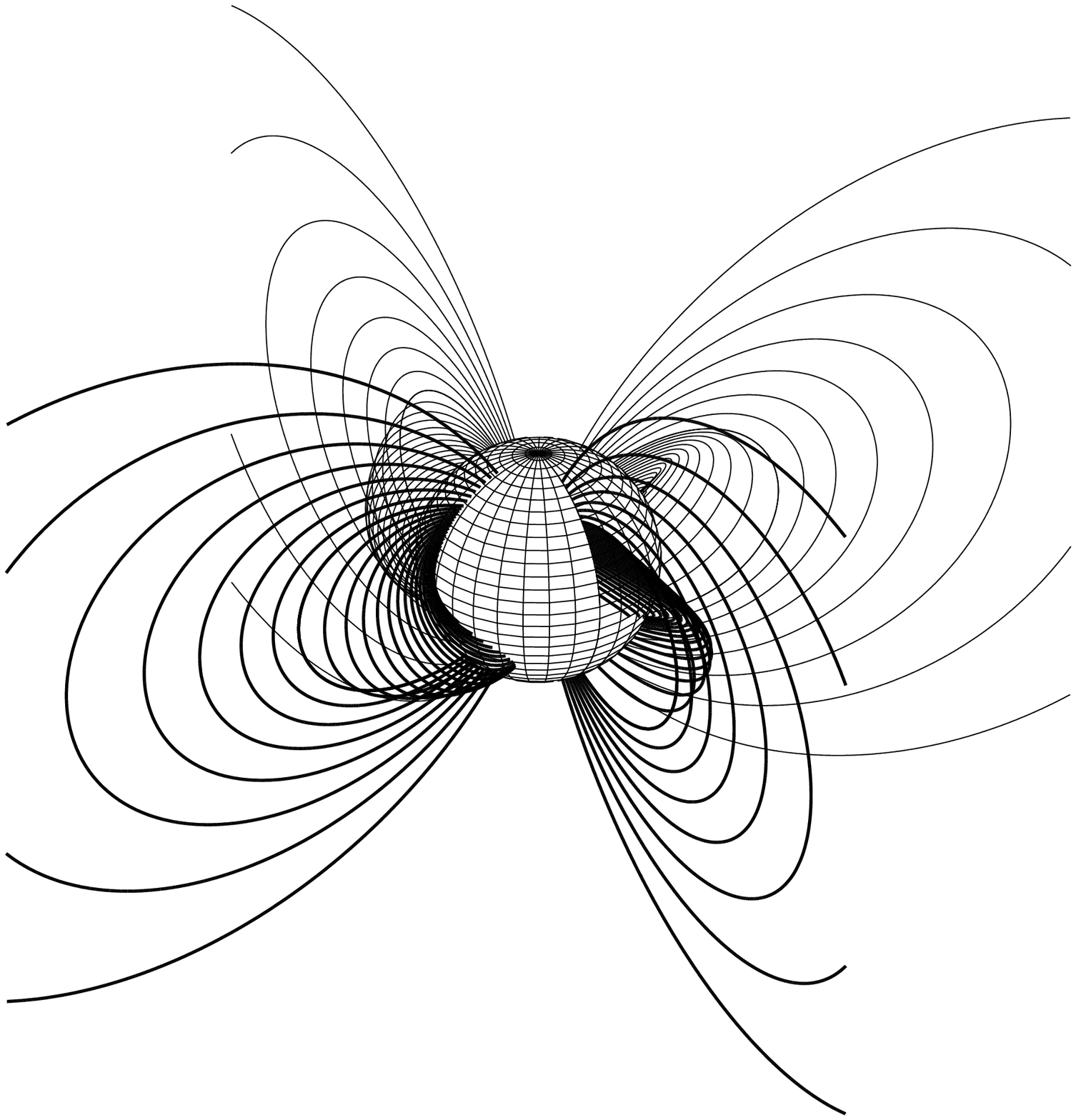}
			\includegraphics[width=0.33\textwidth]{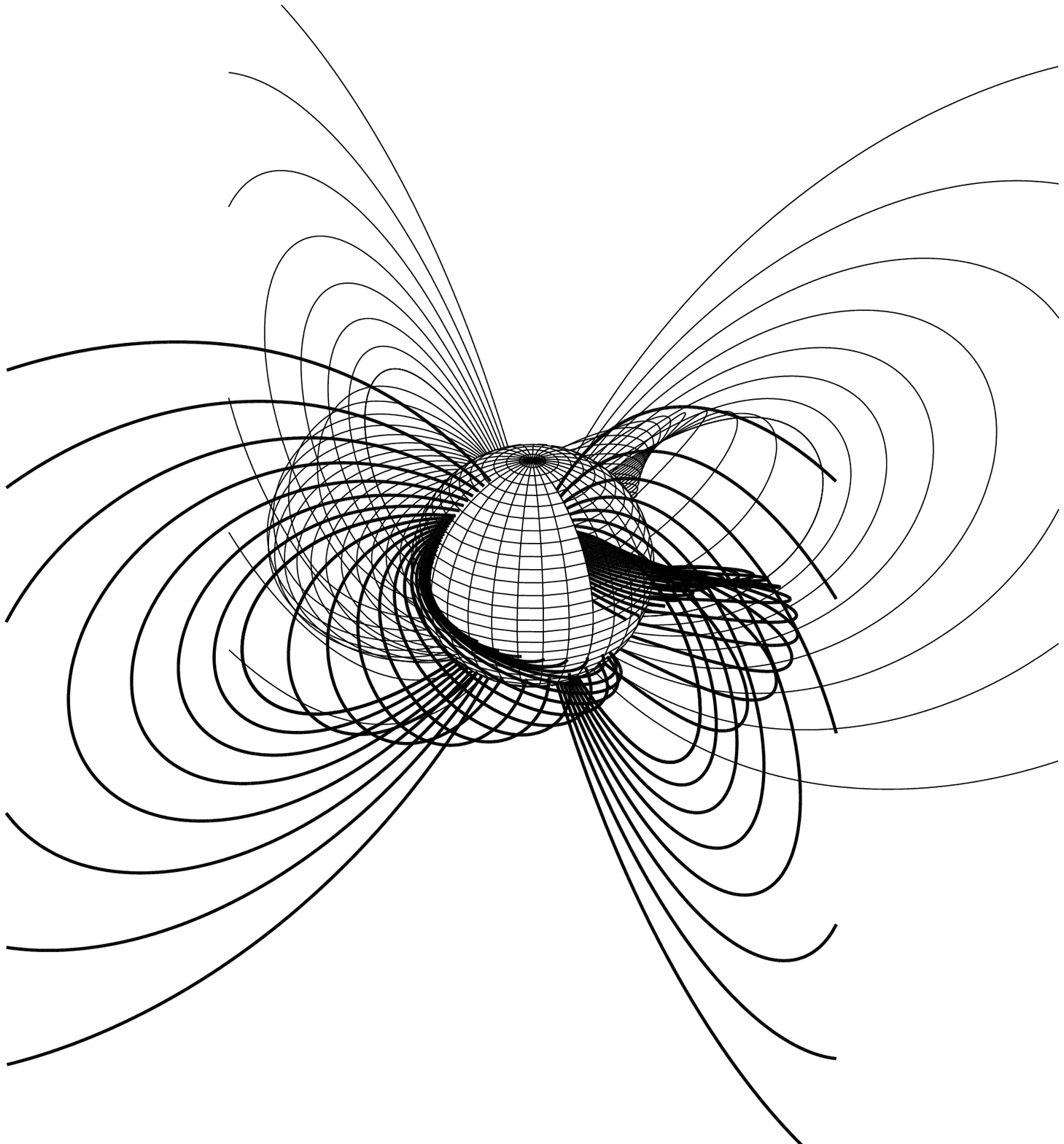}}
		\caption{Left panel: snapshots of the projection of field lines in the meridional plane for model A at three different times: the initial field configuration at $t=0$ is shown with dotted lines; the field at $t=1500$\,yr is shown with dot--dashed lines; and the field at $t=1579$\,yr (the last time step before the magnetosphere saturates) is shown in solid lines. The star is shown in gray, and the distances are rescaled by the stellar radius. The critical field line at $t=1579$\,yr (confining the magnetospheric currents) is shown thicker. Note the accelerating expansion of the field lines over the last two snapshots. Middle and right panels: 3D field lines for the initial magnetic field (at $t=0$) and for the last snapshot before reaching the critical point (at $t=1579$\,yr), respectively.}
		\label{fig_snapshot}
	\end{figure*}
	
	\begin{figure}
		\centerline{\includegraphics[width=0.45\textwidth]{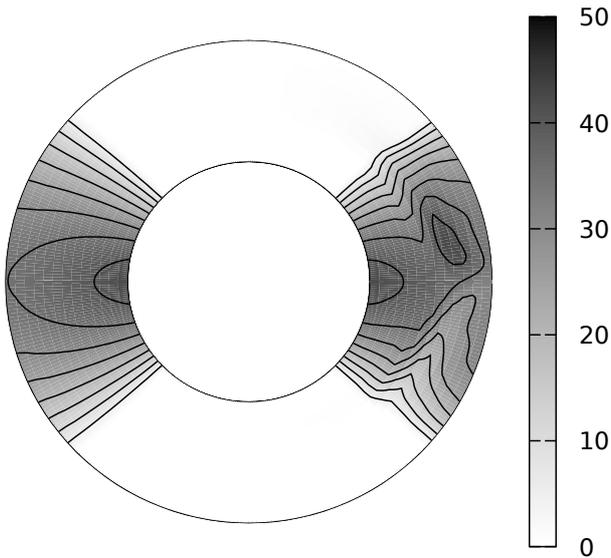}}
		\caption{The toroidal field amplitude $B_\phi$ (gray-scale and isocontours, in units of $10^{12}$\,G) in the crust at the start of the simulation at $t=0$ (left hemisphere) and at $t=1579$\,yr (right hemisphere). The crust extends from the core at $R_{\rm core} = 10.8$\,km to the surface at $R_\star = 11.6$\,km, and it is shown here linearly stretched in the radial direction for visualization purposes.}
		\label{fig_snapshot_crust}
	\end{figure}

\section{Results}\label{section_results}
We next present results from various runs for the magnetosphere, as it adapts to the internal evolution. In the leftmost panel of Fig.~\ref{fig_snapshot}, we show the 2D projection of the magnetic field lines (on the $r\theta$ plane) at several snapshots throughout the evolution of the sample field defined as model A in \S\ref{section_initial}. In this case, the initial field is dipolar, with a toroidal component of relative strength $\mathfrak{f} = 0.6$, confined within the critical field line defined by $P_{\rm c} = 0.5 P_{\rm o}$. Crucially, a value of $P_{\rm c} < P_{\rm o}$ allows for a toroidal field to be present in the magnetosphere at the start of our simulation.

We find that, starting with an initial toroidal field occupying part of the magnetosphere, it gradually increases in amplitude over the course of the evolution. Consequently, the strength of magnetospheric currents, total energy, helicity and twist in the magnetosphere steadily increase over time, and the field lines stretch outwards as the simulation progresses. This continues until the toroidal field reaches a maximal amplitude, beyond which simple field configurations with lines connected to the interior are no longer feasible solutions of the Grad--Shafranov equation (see Paper I), and thus the magnetosphere reaches what we refer to as a \emph{critical point}. For the particular model presented here, this happens at approximately $t_{\rm max} \approx 1579$ yr. It is important to notice that $t_{\rm max}$ is not very sensitive to the exact twist at which the reconnection event occurs because, once the twist is significant ($\sim 1$\,rad), the evolution proceeds very rapidly. At this moment, we expect the magnetosphere to undergo a violent rearrangement on the Alfv\'{e}n timescale, resulting in the expulsion of some or all of the excess energy stored in the toroidal component. After this rearrangement, the evolution will continue, starting from the new initial state, and the process is repeated. Thus, one would expect recurrent outbursts or flares  throughout the evolution, each resulting in the expulsion of a fraction of the available energy. The amount of energy available for such bursts is set by the amplitude of the internal magnetic field, and the recurrence time of the bursts roughly depends on the ratio of the toroidal and poloidal components.

The 3D magnetospheric field configurations at the start (at $t=0$) and end (at $t=1579$\,yr) of the simulation are depicted in the middle and right panels of Fig.~\ref{fig_snapshot}, where now the increase in the twist between the two cases is apparent.

The magnetic field evolution is driven by the Hall and Ohmic terms acting in the crust, which on these short timescales have minimal effect on the poloidal field structure of the interior, while the core field in these models does not evolve. (We do not consider ambipolar diffusion in this paper.) However, the 
apparently small changes of the subsurface magnetic field modify the inner boundary condition for the magnetospheric solution, resulting in a noticeable change of the magnetospheric field lines. We note that increasing the field strength would speed up the evolution because the dominant term in this regime (Hall drift) has a characteristic timescale that decreases linearly with the field strength \citep{1992ApJ...395..250G}.

Although the poloidal component of the internal magnetic field is not seen to substantially change, the toroidal field starts developing asymmetric  quadrupolar structures within the crust (since a dipolar poloidal field gives rise to a quadrupolar toroidal field through the Hall term). This is shown in Fig.~\ref{fig_snapshot_crust}, which is a radially stretched plot of the toroidal field amplitude of the crust at $t=0$ (left hemisphere) and $t=1579$\,yr (right hemisphere). A detailed comparison of the interior (crustal) field evolution for models with vacuum vs.\ force-free magnetospheres is left for a forthcoming paper.

\subsection{Evolution as a function of the initial toroidal amplitude}
The toroidal field strength, magnetic energy, helicity and twist all seem to increase in the exterior during the evolution. The rate of increase depends monotonically on the initial ratio of the toroidal field with respect to the poloidal field (the fraction $\mathfrak{f}$ defined in equation \ref{fraction}). For weak toroidal fields (with $\mathfrak{f} < 0.1$ for the models considered here) the magnetosphere is maintained on long timescales ($\sim 100$\,kyr). For stronger toroidal fields, the region containing currents gradually enlarges with time, until a certain point when no more force-free solutions can be found. The stronger the field, the shorter the time to reach this critical point. Once the magnetosphere reaches this point, the magnetic field must undergo some global reorganization, releasing part of the energy stored in the magnetosphere: the difference between the actual magnetospheric energy ${\cal E}$ and that of the lowest energy model ${\cal E}_{\rm vac}$ (the vacuum solution) sets an upper limit on the available energy budget for rapid magnetospheric events (i.e.\ ${\cal E - E}_{\rm vac}$). In particular, for a magnetic field of a given strength at the pole (at the surface), the external vacuum field configuration with the lowest energy is that of a dipole, which is
	\beq
	{\cal E}_{\rm vac} = \frac{B_{\rm pole}^2 R_\star^3}{12}
	\approx 1.30 \times 10^{45} \, \rm{erg.}
	\label{vacuum}
	\enq
Here we have taken the stellar radius as $R_\star = 11.6$\,km, and the field amplitude at the pole is $B_{\rm pole}=10^{14}$\,G. Numerical simulations show that $\sim 50$ to $70$\% of the available energy may be either dissipated or expelled during this reconnection event \citep{2013ApJ...774...92P}.

	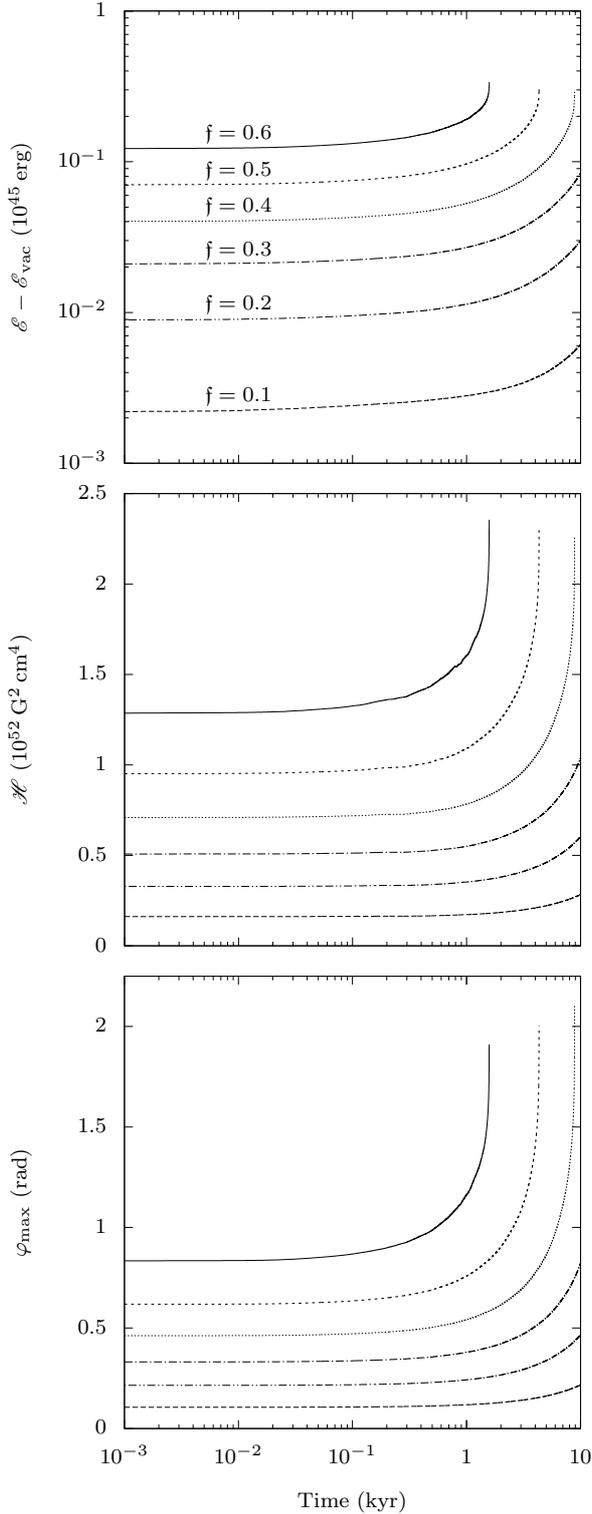
\begin{figure}
	\centering
	\input{figures/epslatex/threeplots}
	\caption{From top to bottom: evolution of the magnetic energy ${\cal E}$ (expressed as the difference with the vacuum dipole, ${\cal E-E}_{\rm vac}$, see equation \ref{vacuum}), helicity ${\cal H}$ and maximum twist $\varphi_{\rm max}$ in the magnetosphere for a model with $P_{\rm c}=0.5 P_{\rm o}$ and for various toroidal field amplitudes, expressed through the fraction $\mathfrak{f}$ defined in equation (\ref{fraction}). All quantities (${\cal E}$, ${\cal H}$ and $\varphi_{\rm max}$) are monotonic functions of $\mathfrak{f}$.}
	\label{fig_energy}
	\end{figure}

In Fig.~\ref{fig_energy}, we show the evolution of the magnetic energy ${\cal E}$ stored in the entire magnetosphere (from the stellar surface up to infinity) for a model with $P_{\rm c}=0.5 P_{\rm o}$ and for various toroidal field amplitudes. We note that the maximum energy of the twisted magnetosphere is about $\sim 30\%$ larger than the corresponding vacuum configuration, although in some cases the available energy budget may be even larger.\footnote{Magnetospheres with disconnected domains can store somewhat more energy (perhaps as much as $\sim 50\%$ larger than the vacuum energy), but these configurations represent degeneracies in the solution of the Grad--Shafranov equation and it is not clear how they could form, as noted in Paper I.} For a $10^{14}$\,G field at the pole, this implies that up to $\sim 4\times 10^{44}$\,erg can be released as a consequence of a large-scale magnetic field reconfiguration.

For $\mathfrak{f} = 0.6$ (corresponding to the uppermost solid line in the plots in Fig.~\ref{fig_energy}), the power gain in the magnetosphere is of the order of several $\sim 10^{33}$\,erg/s, or about $\sim 10^{41}$\,erg/yr. In other words, the amount of energy stored into the magnetosphere (from the interior) over the course of a year is about four orders of magnitude less than the total energy of the magnetosphere. This, incidentally, is also about the accuracy in the calculation of the energy (as noted in Paper I), and energy conservation taking into account the energy loss in the interior (due to Joule heating and Poynting flux) and the energy gain in the exterior is not satisfied better than at this level.

The plots for the helicity ${\cal H}$ and maximum twist $\varphi_{\rm max}$ show that the initial values of both quantities can be significantly increased (by factors of up to $3$--$4$) over the course of evolution. However, note that for all models shown here, the maximum helicity achieved in the first $10$\,kyr is $\lesssim 2.5 \times 10^{52}$\,G$^2$\,cm$^4$. The maximum twist, on the other hand, is capped at $\sim 2$\,rad for these models.

	\begin{figure}
	\centering
	\input{figures/epslatex/data}
	\caption{Maximum time $t_{\rm max}$ (time to reach the critical point) versus the fraction $\mathfrak{f}$ of toroidal to poloidal amplitude (equation \ref{fraction}), for $P_{\rm c} = 0.5 P_{\rm o}$, and for a number of models with $0.35 \leqslant \mathfrak{f} \leqslant 0.7$, some of which are plotted in Fig.~\ref{fig_energy}. The limiting value $\mathfrak{f}_{\rm lim} = 0.701$ is determined numerically. Above that value, solutions for the magnetosphere cannot be constructed even at the start of the simulation.}
	\label{fig_data}
	\end{figure}
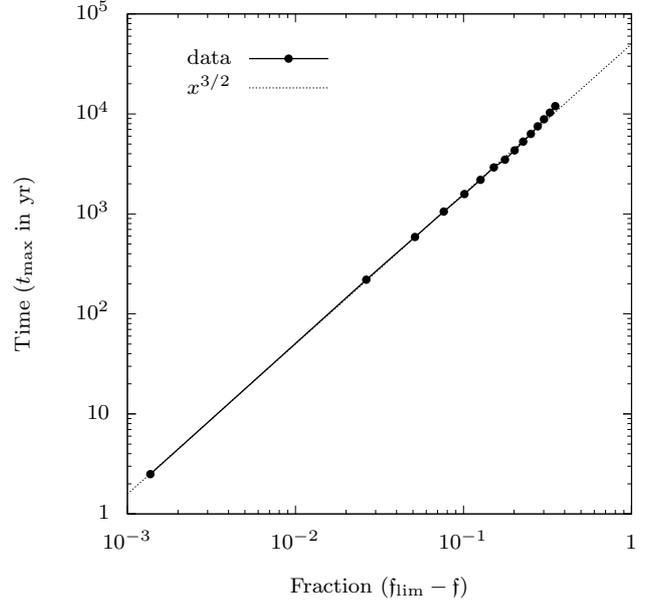

The timescale to reach the critical point ($t_{\rm max}$) depends monotonically on the ratio of the toroidal field to the poloidal field (given through the fraction $\mathfrak{f}$ defined in equation \ref{fraction}), as can be seen in Fig.~\ref{fig_data}. On a log--log plot, it becomes apparent that the two quantities are related by a power-law of the form
	\beq
	t_{\rm max} \approx 5\times 10^4 (\mathfrak{f}_{\rm lim} - \mathfrak{f})^{3/2} \, \rm{yr}.
	\enq
The value of $\mathfrak{f}_{\rm lim}=0.701$ for this model has been determined by fitting the data points, and sets the upper limit for the toroidal field. For larger values, no solution of the Grad--Shafranov equation can be constructed even at the start of the simulation.

After the magnetospheric rearrangement, a new magnetosphere with lower helicity content and smaller twist will be formed, and the slow, internally driven evolution will continue. In principle, the scalability of the amplitudes for the magnetosphere implies that, by a suitable choice of the amplitudes of the poloidal and toroidal fields in the interior, within the range of usual NS field strengths, any desirable amount of energy in the range $10^{42}-10^{45}$\,erg (typical of magnetar outburst/flare energetics) can be achieved. The recurrence timescale may range from a few decades up to several millennia (or more).

\subsection{Profiles of the toroidal function, twist and voltage throughout the evolution}
\begin{figure}
	\centering
	\input{figures/epslatex/voltage}
	\caption{From top to bottom: toroidal function $T$, twist $\varphi$, surface velocity of footprints $v_\varphi$ and voltage $V$ as functions of the surface poloidal function $P$ at several snapshots throughout the evolution for model A (depicted in Fig.~\ref{fig_snapshot}). Here $P$ is shown normalized by its largest value at the surface $P_{\rm o}$, which itself varies in time (together with the critical value $P_{\rm c}$). $T$ is then given in units of $P_{\rm o}/R_\star$ where $R_\star$ is the stellar radius. $\varphi$ is given in radians, $v_\varphi$ in km/yr and $V$ in megavolts (MV). $V$ is integrated back from $P=P_{\rm o}$ (see equation \ref{voltage}) and is therefore constant below $P_{\rm c}$.}
	\label{fig_voltage}
	\end{figure}
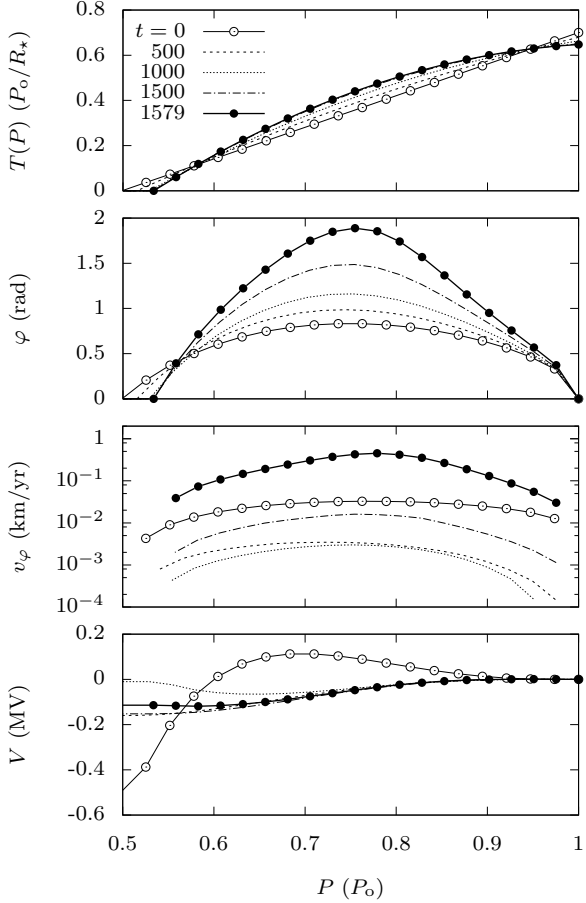

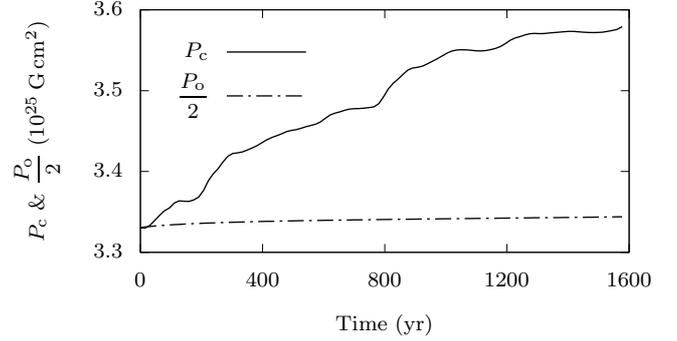
\begin{figure}
	\centering
	\input{figures/epslatex/pcritical}
	\caption{Evolution of $P_{\rm c}$ and $P_{\rm o}$ for model A. $P_{\rm c}$ defines the magnetic surface encompassing the magnetospheric currents, and $P_{\rm o}$ is the maximum value of the poloidal stream function at the surface. Initially $P_{\rm c}$ is set to $P_{\rm o}/2$, but over time it slightly increases by up to $\sim 10\%$. The increase in $P_{\rm o}$ in the same time interval is much less noticeable ($\lesssim 1\%$).}
	\label{fig_pcritical}
	\end{figure}

In Fig.~\ref{fig_voltage}, we show the evolution of various quantities as functions of $P$ (which itself can be read as some function of latitude) for model A ($P_{\rm c} = 0.5 P_{\rm o}$ and $\mathfrak{f}=0.6$). The top panel shows the profile of the function $T(P)$, which remains fairly stable throughout the evolution, at five instances. (The lines for $t=1500$\,yr and $t=1579$\,yr are indistinguishable on the scale of the plot.) The poloidal function $P$ is expressed in units of $P_{\rm o}$, and the toroidal function $T$ is given in units of $P_{\rm o}/R_\star$. Both $P_{\rm o}$ and $P_{\rm c}$ are functions of time (shown in Fig.~\ref{fig_pcritical}). The horizontal scale only shows the region $P > 0.5 P_{\rm o}$, where currents exist.

The second panel from the top shows the twist of the different lines. At each instant, the twist is zero at $P_{\rm c}$ where the toroidal field vanishes, and at $P_{\rm o}$ where the field line length goes to zero. In the figure, the twist can be seen to gradually increase over the course of the evolution, until it reaches the maximum value of $\sim 2$\,rad. The third panel shows the surface velocity of the footprints (of the field lines) defined as
	\beq 
	v_\varphi \equiv \dot{\varphi} ~r \sin\theta \, .
	\label{vphi}
	\enq
Following the rapid transient stage (apparent at $t=0$) which appears as a consequence of the relaxation of the initial magnetic field configuration, this velocity is maintained at relatively small values, of the order of $10^{-3}$\,km/yr for much of the evolution (for $t=500$\,yr and $t=1000$\,yr). In the late stages of evolution, close to the critical point (for $t=1500$\,yr and $t=1579$\,yr), the velocity increases and can reach relatively high values, of the order of 1\,km/yr.

Finally, using the formalism developed by \cite{2009ApJ...703.1044B}, we estimate the inductive voltage $V$ between the footprints of a twisted field line in the magnetosphere (see his equation 17), which in our notation can be written as
	\beq
	\left(\frac{\partial V}{\partial P}\right)_t = \frac{\varphi}{2\pi c}
	\left(\frac{1}{T}\frac{\partial T}{\partial t} \right)_P \, .
	\label{voltage}
	\enq
Bear in mind that this equation is derived under a number of very specific assumptions (e.g.\ small twist), and therefore only serves as a rather rough estimate. In our case, the field lines are not particularly close to those of a vacuum dipole, but are rather notably different, especially near the critical point, as can be seen in Fig.~\ref{fig_snapshot}. Moreover, the strong twist significantly inclines the field line plane (as is apparent from the 3D field structure shown in Fig.~\ref{fig_snapshot}), which is also not accounted for in the derivation of the above equation.

In this paper, we do not consider the relatively fast outburst decay, but rather the slow process during which the magnetosphere is replenished with plasma and the twist increases until some instability triggers an outburst or a flare. The strong observational evidence for the presence of hard tails in the spectra of magnetars in quiescence \citep{2008ApJ...686.1245R}, long before an outburst occurs, indicates that there must be two different magnetospheric regimes: one stable, slow process (on the scale of tens to hundreds of years) to increase the currents and twist (this paper), and a fast dissipation on a timescale of months/a few years, once some critical point is reached. In the stable regime, the results shown in the bottom panel of Fig.~\ref{fig_voltage} indicate that, during the epoch of quasi-static increase of twist on long timescales, the voltage along a field line is moderate, of the order of $0.1$\,MV.\footnote{Here, we choose to express the voltage in the SI units of \emph{volts}, rather than the cgs units of \emph{statvolts}.} This voltage is well below the critical value needed to initiate a discharge ($\sim$\,GV) that is supposed to be reached during the outburst peak and subsequent decay in the untwisting magnetar model \citep{2009ApJ...703.1044B}. Beyond this upper critical limit, pair creation is activated and the voltage will oscillate around some average value in the $\sim$\,GV range on much shorter timescales \citep{2007ApJ...657..967B}.

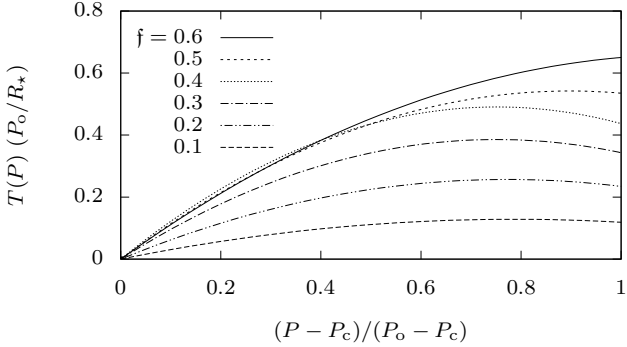
\begin{figure}
	\centering
	\input{figures/epslatex/t_profile}
	\caption{Sample profiles of the toroidal function $T(P)$ (in units of $P_{\rm o}/R_\star$) for the same six models shown in Fig.~\ref{fig_energy}, at various instances in their evolution. (For $\mathfrak{f} = 0.1$, $0.2$ and $0.3$ the profiles are shown at $t = 10$\,kyr, while for the remaining cases they are shown close to their respective critical points: for $\mathfrak{f} = 0.4$ at $t = 8.5$\,kyr, for $\mathfrak{f} = 0.5$ at $t = 4$\,kyr, and for $\mathfrak{f} = 0.6$ at $t = 1.5$\,kyr.) Since $P_{\rm c}$ and $P_{\rm o}$ are slightly different for each model, we renormalize $P$ as indicated in the horizontal axis label. (The toroidal field is present only in the interval $P_{\rm c} \leqslant P \leqslant P_{\rm o}$.)}
	\label{fig_t_profile}
	\end{figure}

As noted in \S\ref{section_boundary}, the function $T(P)$ (of the form given by equation \ref{Tfit}) may have a maximum somewhere in the interval $P_{\rm c} < P < P_{\rm o}$, implying that the current (given by equation \ref{current}) will go through zero at some magnetic surface and then reverse direction. Indeed, for weaker toroidal fields ($\mathfrak{f} \lesssim 0.5$), we do obtain such configurations at later stages in the evolution, as shown in Fig.~\ref{fig_t_profile}. Such models will be explored in more detail in future work.

\subsection{Spin-down properties}
The loss of rotational energy can be estimated in terms of a simple rotating oblique dipole. We calculate the period derivative (spindown rate) through the relation \citep{2006ApJ...648L..51S}
	\beq
	{\cal P} {\cal \dot{P}} = \frac{\pi^2 R_\star^6}{I c^3} {\cal B}^2 f_{\chi} \, .
	\label{ppdot}
	\enq
(We use the calligraphic letter ${\cal P}$ for the period in order to avoid confusion with the poloidal function $P$.) Here ${\cal B}$ is the amplitude of the \emph{dipolar} component of the poloidal field at the pole. For a force-free twisted magnetosphere, the effective dipolar component is larger than for the vacuum case, and as a result the period derivative is larger. The factor $f_\chi$ is of the form
	\beq
	f_\chi = \kappa_0 + \kappa_1 \sin^2\chi \, ,
	\enq
where $\chi$ is the angle of inclination between the magnetic and rotation axes. For a vacuum magnetsphere $\kappa_0 = 0$ and $\kappa_1= 2/3$, and for a force-free twisted magnetosphere $\kappa_0 \approx \kappa_1 \approx 1$ \citep{2014MNRAS.441.1879P}. We take the average value $f_\chi = 3/2$ in our calculations. For the stellar model used in this work ($M_\star = 1.4 M_\odot$ and $R_\star = 11.6$\,km) the moment of inertia is $I \approx 1.62\times 10^{45}$\,g\,cm$^2$.

	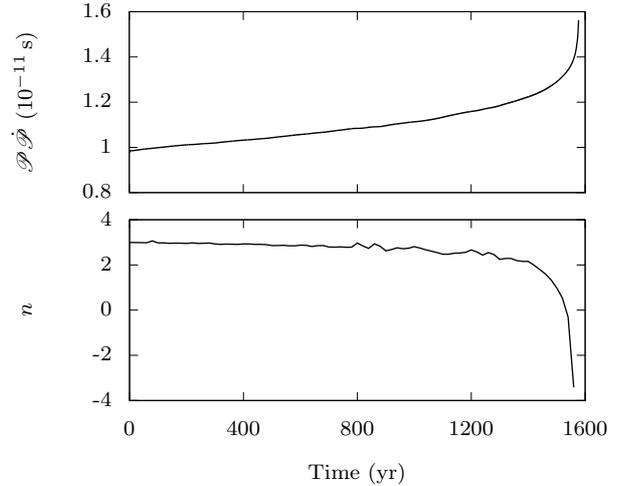
\begin{figure}
	\centering
	\input{figures/epslatex/spindown}
	\caption{The product of the period $\cal{P}$ and period derivative $\cal{\dot{P}}$ (equation \ref{ppdot}), and braking index $n$ (equation \ref{braking}) as functions of time for model A.}
	\label{fig_spindown}
	\end{figure}

The product of the period ${\cal P}$ and the period derivative ${\cal \dot{P}}$ for model A is shown in the upper panel of Fig.~\ref{fig_spindown}, in units of $10^{-11}$\,s. For typical magnetar periods of ${\cal P} \sim 10$\,s, this gives ${\cal \dot{P}} \sim 10^{-12}$, which is consistent with observations \citep{2013MNRAS.434..123V}. Note that the plot for ${\cal P \dot{P}}$ can be read as a plot of ${\cal B}^2$ through equation (\ref{ppdot}). The gradual increase in ${\cal B}$ over the course of the evolution translates into an increase in ${\cal \dot{P}}$ by a corresponding amount, in this case up to $\sim 60\%$ with respect to the vacuum dipole model (at $t=0$), as can be seen in the figure.

In the lower panel of Fig.~\ref{fig_spindown}, we show the braking index $n$, which is defined through
	\beq
	n = \frac{\Omega \ddot{\Omega}}{~\dot{\Omega}^2}
	= 2 - \frac{{\cal P \ddot{P}}}{{\cal \dot{P}}^2}
	= 3 - 2 \frac{{\cal \dot{B} P}}{{\cal B \dot{P}}} \, .
	\label{braking}
	\enq
In the plot, we take ${\cal P}_0 = 10$\,ms for the initial period. Thus, here we are considering the time shortly after the formation of a magnetar, up to the first potential magnetospheric flare. The spindown at this stage is very rapid (because ${\cal \dot{P}} \propto {\cal P}^{-1}$, from equation \ref{ppdot}) and the period quickly increases to about $1$\,s by the time the magnetosphere reaches the critical point. In this early phase, the braking index is in the vicinity of 3, but gradually decreasing. Starting with a higher initial period would effectively displace the plot for $n$ slightly downwards (while also increasing numerical noise due to the calculation of the time derivatives of ${\cal B}$ and ${\cal P}$).

The braking indices of magnetars cannot be determined accurately, although a few estimates have been found to lie within $1 < n < 3$ \citep{2016MNRAS.456...55G}, and recently \cite{2017ApJ...843L...1L} inferred $n \sim 2.6$ and $\sim 2.9$ for two objects believed to be millisecond magnetars. These values are comparable to those of the few actually measured braking indices of pulsars \citep{2015MNRAS.446..857L}. This range is also in agreement with our results shown in Fig.~\ref{fig_spindown}, except for the last few decades immediately before the critical point, where the braking index dramatically drops into negative values. Such a sudden drop would be expected to happen when the surface magnetic field strength is increasing in an accelerating manner near the critical point. An unusually low braking index (below 1) could therefore (with certain caution) be taken as a potential indication of an imminent magnetospheric event, probably within a few months or years.

\section{Conclusions}\label{section_conclusions}
We have studied the magnetospheric evolution of magnetars coupled to their long-term internal evolution. We have shown how the continuous transfer of energy and helicity from the NS crust, driven by Ohmic dissipation and Hall drift, results in long-lived magnetospheric currents that can explain the spectral features associated with magnetars. We find that the magnetospheric twist grows continuously until a critical point is reached, where some instabilities may arise. The maximum twist attained is $\varphi_{\rm max} \sim 2$\,rad, in agreement with other estimates from previous works obtained through a variety of methods \citep{1994ApJ...430..898M,2002ApJ...574..332T,2012ApJ...754L..12P,2013ApJ...774...92P,2017MNRAS.468.2011K}. For poloidal and toroidal field strengths in the range of $10^{14}$\,G, we show that the energy available for outbursts and flares can be up to a few $10^{44}$\,erg. Increasing the magnetic field strength by 10 would increase this energy budget by a factor of 100. The typical voltages along magnetic field lines during this quasi-steady evolution are much weaker than the critical voltage used to untwist the magnetosphere on timescales of years during the outburst decay, as discussed in \cite{2007ApJ...657..967B,2009ApJ...703.1044B,2012ApJ...754L..12P}. 

Thus, our mechanism, driven by the internal magnetic field evolution, explains how the twist builds up gradually, displacing footprints at an increasing rate that can rise up to $1$\,km/yr in the very final stages of evolution, prior to reaching the critical point when a global current rearrangement must occur. If reconnection resulted in the emission of a large plasmoid, we would be in the scenario of a magnetar flare, while potentially strong returning currents in combination with internal heating can create long-lived hot spots such as those inferred from observations during the radiative outburst decay \citep{2015MNRAS.452.3357G}. The excess energy stored is then freed, and the process would start over again from a new magnetosphere solution with lower twist and helicity. Due to the very different timescales, we cannot follow in detail the last period of the magnetospheric inflation, and its dynamical reorganization, which must be done dynamically. During this last epoch, before instabilities set in, the voltage along field lines should grow up to the GV range, but how this happens remains to be studied in detail.

We will describe how the magnetosphere affects the interior in future work, where we will analyze the results of our simulations focusing on the evolution of the crust and the subsurface layers. In particular, we will address issues about the effects of higher order multipoles, important in the vicinity of the stellar surface, and the role of elasticity in the crust.

%% file: figures/epslatex/threeplots.tex
\begingroup
  \makeatletter
  \providecommand\color[2][]{%
    \GenericError{(gnuplot) \space\space\space\@spaces}{%
      Package color not loaded in conjunction with
      terminal option `colourtext'%
    }{See the gnuplot documentation for explanation.%
    }{Either use 'blacktext' in gnuplot or load the package
      color.sty in LaTeX.}%
    \renewcommand\color[2][]{}%
  }%
  \providecommand\includegraphics[2][]{%
    \GenericError{(gnuplot) \space\space\space\@spaces}{%
      Package graphicx or graphics not loaded%
    }{See the gnuplot documentation for explanation.%
    }{The gnuplot epslatex terminal needs graphicx.sty or graphics.sty.}%
    \renewcommand\includegraphics[2][]{}%
  }%
  \providecommand\rotatebox[2]{#2}%
  \@ifundefined{ifGPcolor}{%
    \newif\ifGPcolor
    \GPcolorfalse
  }{}%
  \@ifundefined{ifGPblacktext}{%
    \newif\ifGPblacktext
    \GPblacktexttrue
  }{}%
  \let\gplgaddtomacro\g@addto@macro
  \gdef\gplbacktext{}%
  \gdef\gplfronttext{}%
  \makeatother
  \ifGPblacktext
    \def\colorrgb#1{}%
    \def\colorgray#1{}%
  \else
    \ifGPcolor
      \def\colorrgb#1{\color[rgb]{#1}}%
      \def\colorgray#1{\color[gray]{#1}}%
      \expandafter\def\csname LTw\endcsname{\color{white}}%
      \expandafter\def\csname LTb\endcsname{\color{black}}%
      \expandafter\def\csname LTa\endcsname{\color{black}}%
      \expandafter\def\csname LT0\endcsname{\color[rgb]{1,0,0}}%
      \expandafter\def\csname LT1\endcsname{\color[rgb]{0,1,0}}%
      \expandafter\def\csname LT2\endcsname{\color[rgb]{0,0,1}}%
      \expandafter\def\csname LT3\endcsname{\color[rgb]{1,0,1}}%
      \expandafter\def\csname LT4\endcsname{\color[rgb]{0,1,1}}%
      \expandafter\def\csname LT5\endcsname{\color[rgb]{1,1,0}}%
      \expandafter\def\csname LT6\endcsname{\color[rgb]{0,0,0}}%
      \expandafter\def\csname LT7\endcsname{\color[rgb]{1,0.3,0}}%
      \expandafter\def\csname LT8\endcsname{\color[rgb]{0.5,0.5,0.5}}%
    \else
      \def\colorrgb#1{\color{black}}%
      \def\colorgray#1{\color[gray]{#1}}%
      \expandafter\def\csname LTw\endcsname{\color{white}}%
      \expandafter\def\csname LTb\endcsname{\color{black}}%
      \expandafter\def\csname LTa\endcsname{\color{black}}%
      \expandafter\def\csname LT0\endcsname{\color{black}}%
      \expandafter\def\csname LT1\endcsname{\color{black}}%
      \expandafter\def\csname LT2\endcsname{\color{black}}%
      \expandafter\def\csname LT3\endcsname{\color{black}}%
      \expandafter\def\csname LT4\endcsname{\color{black}}%
      \expandafter\def\csname LT5\endcsname{\color{black}}%
      \expandafter\def\csname LT6\endcsname{\color{black}}%
      \expandafter\def\csname LT7\endcsname{\color{black}}%
      \expandafter\def\csname LT8\endcsname{\color{black}}%
    \fi
  \fi
    \setlength{\unitlength}{0.0500bp}%
    \ifx\gptboxheight\undefined%
      \newlength{\gptboxheight}%
      \newlength{\gptboxwidth}%
      \newsavebox{\gptboxtext}%
    \fi%
    \setlength{\fboxrule}{0.5pt}%
    \setlength{\fboxsep}{1pt}%
\begin{picture}(5102.00,11338.00)%
    \gplgaddtomacro\gplbacktext{%
      \csname LTb\endcsname%
      \put(1058,11223){\makebox(0,0)[r]{\strut{}$1$}}%
      \put(1058,7823){\makebox(0,0)[r]{\strut{}$10^{-3}$}}%
      \put(1058,8956){\makebox(0,0)[r]{\strut{}$10^{-2}$}}%
      \put(1058,10090){\makebox(0,0)[r]{\strut{}$10^{-1}$}}%
      \put(1190,7603){\makebox(0,0){\strut{}}}%
      \put(2040,7603){\makebox(0,0){\strut{}}}%
      \put(2890,7603){\makebox(0,0){\strut{}}}%
      \put(3740,7603){\makebox(0,0){\strut{}}}%
      \put(4590,7603){\makebox(0,0){\strut{}}}%
      \put(2040,8330){\makebox(0,0){\strut{}$\mathfrak{f}=0.1$}}%
      \put(2040,9021){\makebox(0,0){\strut{}$\mathfrak{f}=0.2$}}%
      \put(2040,9427){\makebox(0,0){\strut{}$\mathfrak{f}=0.3$}}%
      \put(2040,9758){\makebox(0,0){\strut{}$\mathfrak{f}=0.4$}}%
      \put(2040,10024){\makebox(0,0){\strut{}$\mathfrak{f}=0.5$}}%
      \put(2040,10305){\makebox(0,0){\strut{}$\mathfrak{f}=0.6$}}%
    }%
    \gplgaddtomacro\gplfronttext{%
      \csname LTb\endcsname%
      \put(420,9523){\rotatebox{-270}{\makebox(0,0){\strut{}${\cal E-E}_{\rm vac}$ ($10^{45}$\,erg)}}}%
    }%
    \gplgaddtomacro\gplbacktext{%
      \csname LTb\endcsname%
      \put(1058,4195){\makebox(0,0)[r]{\strut{}$0$}}%
      \put(1058,4875){\makebox(0,0)[r]{\strut{}$0.5$}}%
      \put(1058,5555){\makebox(0,0)[r]{\strut{}$1$}}%
      \put(1058,6235){\makebox(0,0)[r]{\strut{}$1.5$}}%
      \put(1058,6915){\makebox(0,0)[r]{\strut{}$2$}}%
      \put(1058,7595){\makebox(0,0)[r]{\strut{}$2.5$}}%
      \put(1190,3975){\makebox(0,0){\strut{}}}%
      \put(2040,3975){\makebox(0,0){\strut{}}}%
      \put(2890,3975){\makebox(0,0){\strut{}}}%
      \put(3740,3975){\makebox(0,0){\strut{}}}%
      \put(4590,3975){\makebox(0,0){\strut{}}}%
    }%
    \gplgaddtomacro\gplfronttext{%
      \csname LTb\endcsname%
      \put(420,5895){\rotatebox{-270}{\makebox(0,0){\strut{}${\cal H}$ ($10^{52}$\,G$^2$\,cm$^4$)}}}%
    }%
    \gplgaddtomacro\gplbacktext{%
      \csname LTb\endcsname%
      \put(1057,566){\makebox(0,0)[r]{\strut{}$0$}}%
      \put(1057,1322){\makebox(0,0)[r]{\strut{}$0.5$}}%
      \put(1057,2078){\makebox(0,0)[r]{\strut{}$1$}}%
      \put(1057,2833){\makebox(0,0)[r]{\strut{}$1.5$}}%
      \put(1057,3589){\makebox(0,0)[r]{\strut{}$2$}}%
      \put(3740,346){\makebox(0,0){\strut{}$1$}}%
      \put(4590,346){\makebox(0,0){\strut{}$10$}}%
      \put(1189,346){\makebox(0,0){\strut{}$10^{-3}$}}%
      \put(2039,346){\makebox(0,0){\strut{}$10^{-2}$}}%
      \put(2889,346){\makebox(0,0){\strut{}$10^{-1}$}}%
    }%
    \gplgaddtomacro\gplfronttext{%
      \csname LTb\endcsname%
      \put(419,2266){\rotatebox{-270}{\makebox(0,0){\strut{}$\varphi_{\rm max}$ (rad)}}}%
      \put(2889,16){\makebox(0,0){\strut{}Time (kyr)}}%
    }%
    \gplbacktext
    \put(0,0){\includegraphics{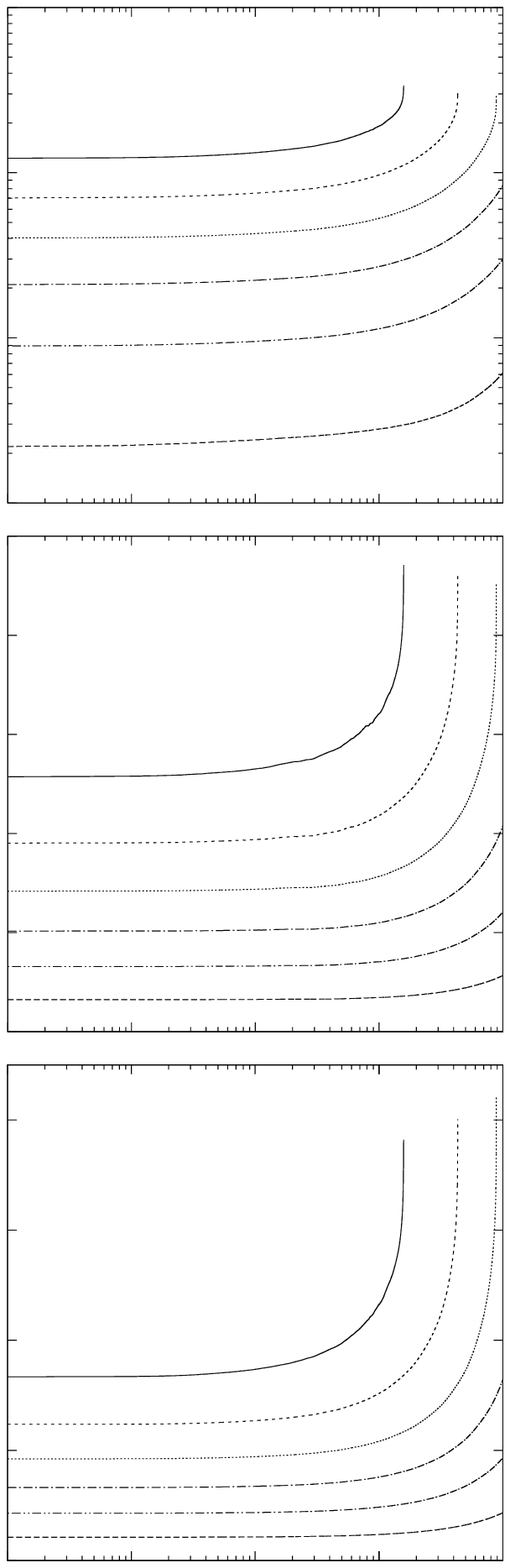}}%
    \gplfronttext
  \end{picture}%
\endgroup

%% file: figures/epslatex/data.tex
\begingroup
  \makeatletter
  \providecommand\color[2][]{%
    \GenericError{(gnuplot) \space\space\space\@spaces}{%
      Package color not loaded in conjunction with
      terminal option `colourtext'%
    }{See the gnuplot documentation for explanation.%
    }{Either use 'blacktext' in gnuplot or load the package
      color.sty in LaTeX.}%
    \renewcommand\color[2][]{}%
  }%
  \providecommand\includegraphics[2][]{%
    \GenericError{(gnuplot) \space\space\space\@spaces}{%
      Package graphicx or graphics not loaded%
    }{See the gnuplot documentation for explanation.%
    }{The gnuplot epslatex terminal needs graphicx.sty or graphics.sty.}%
    \renewcommand\includegraphics[2][]{}%
  }%
  \providecommand\rotatebox[2]{#2}%
  \@ifundefined{ifGPcolor}{%
    \newif\ifGPcolor
    \GPcolorfalse
  }{}%
  \@ifundefined{ifGPblacktext}{%
    \newif\ifGPblacktext
    \GPblacktexttrue
  }{}%
  \let\gplgaddtomacro\g@addto@macro
  \gdef\gplbacktext{}%
  \gdef\gplfronttext{}%
  \makeatother
  \ifGPblacktext
    \def\colorrgb#1{}%
    \def\colorgray#1{}%
  \else
    \ifGPcolor
      \def\colorrgb#1{\color[rgb]{#1}}%
      \def\colorgray#1{\color[gray]{#1}}%
      \expandafter\def\csname LTw\endcsname{\color{white}}%
      \expandafter\def\csname LTb\endcsname{\color{black}}%
      \expandafter\def\csname LTa\endcsname{\color{black}}%
      \expandafter\def\csname LT0\endcsname{\color[rgb]{1,0,0}}%
      \expandafter\def\csname LT1\endcsname{\color[rgb]{0,1,0}}%
      \expandafter\def\csname LT2\endcsname{\color[rgb]{0,0,1}}%
      \expandafter\def\csname LT3\endcsname{\color[rgb]{1,0,1}}%
      \expandafter\def\csname LT4\endcsname{\color[rgb]{0,1,1}}%
      \expandafter\def\csname LT5\endcsname{\color[rgb]{1,1,0}}%
      \expandafter\def\csname LT6\endcsname{\color[rgb]{0,0,0}}%
      \expandafter\def\csname LT7\endcsname{\color[rgb]{1,0.3,0}}%
      \expandafter\def\csname LT8\endcsname{\color[rgb]{0.5,0.5,0.5}}%
    \else
      \def\colorrgb#1{\color{black}}%
      \def\colorgray#1{\color[gray]{#1}}%
      \expandafter\def\csname LTw\endcsname{\color{white}}%
      \expandafter\def\csname LTb\endcsname{\color{black}}%
      \expandafter\def\csname LTa\endcsname{\color{black}}%
      \expandafter\def\csname LT0\endcsname{\color{black}}%
      \expandafter\def\csname LT1\endcsname{\color{black}}%
      \expandafter\def\csname LT2\endcsname{\color{black}}%
      \expandafter\def\csname LT3\endcsname{\color{black}}%
      \expandafter\def\csname LT4\endcsname{\color{black}}%
      \expandafter\def\csname LT5\endcsname{\color{black}}%
      \expandafter\def\csname LT6\endcsname{\color{black}}%
      \expandafter\def\csname LT7\endcsname{\color{black}}%
      \expandafter\def\csname LT8\endcsname{\color{black}}%
    \fi
  \fi
    \setlength{\unitlength}{0.0500bp}%
    \ifx\gptboxheight\undefined%
      \newlength{\gptboxheight}%
      \newlength{\gptboxwidth}%
      \newsavebox{\gptboxtext}%
    \fi%
    \setlength{\fboxrule}{0.5pt}%
    \setlength{\fboxsep}{1pt}%
\begin{picture}(5102.00,5102.00)%
    \gplgaddtomacro\gplbacktext{%
      \csname LTb\endcsname%
      \put(814,891){\makebox(0,0)[r]{\strut{}$1$}}%
      \put(814,1643){\makebox(0,0)[r]{\strut{}$10$}}%
      \put(814,2395){\makebox(0,0)[r]{\strut{}$10^{2}$}}%
      \put(814,3146){\makebox(0,0)[r]{\strut{}$10^{3}$}}%
      \put(814,3898){\makebox(0,0)[r]{\strut{}$10^{4}$}}%
      \put(814,4650){\makebox(0,0)[r]{\strut{}$10^{5}$}}%
      \put(4705,671){\makebox(0,0){\strut{}$1$}}%
      \put(946,671){\makebox(0,0){\strut{}$10^{-3}$}}%
      \put(2199,671){\makebox(0,0){\strut{}$10^{-2}$}}%
      \put(3452,671){\makebox(0,0){\strut{}$10^{-1}$}}%
    }%
    \gplgaddtomacro\gplfronttext{%
      \csname LTb\endcsname%
      \put(176,2770){\rotatebox{-270}{\makebox(0,0){\strut{}Time ($t_{\rm max}$ in yr)}}}%
      \put(2825,341){\makebox(0,0){\strut{}Fraction ($\mathfrak{f}_{\rm lim} - \mathfrak{f}$)}}%
      \put(1721,4314){\makebox(0,0)[r]{\strut{}data}}%
      \put(1721,4094){\makebox(0,0)[r]{\strut{}$x^{3/2}$}}%
    }%
    \gplbacktext
    \put(0,0){\includegraphics{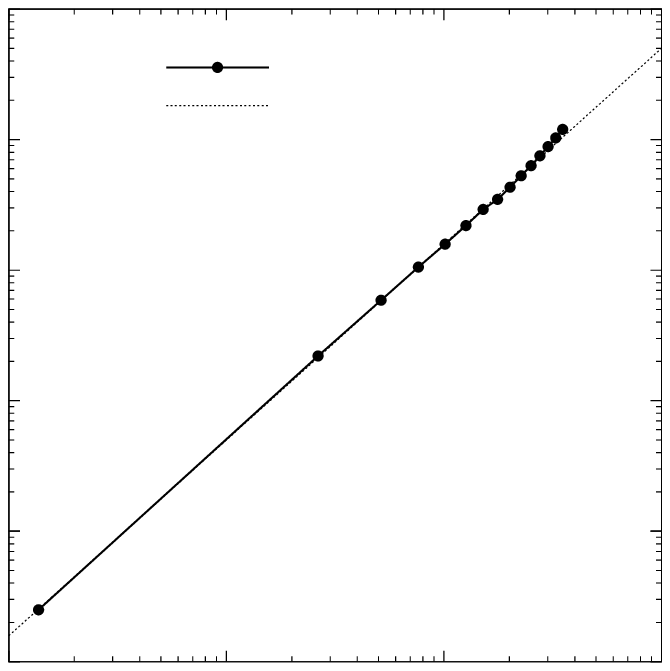}}%
    \gplfronttext
  \end{picture}%
\endgroup

%% file: figures/epslatex/voltage.tex
\begingroup
  \makeatletter
  \providecommand\color[2][]{%
    \GenericError{(gnuplot) \space\space\space\@spaces}{%
      Package color not loaded in conjunction with
      terminal option `colourtext'%
    }{See the gnuplot documentation for explanation.%
    }{Either use 'blacktext' in gnuplot or load the package
      color.sty in LaTeX.}%
    \renewcommand\color[2][]{}%
  }%
  \providecommand\includegraphics[2][]{%
    \GenericError{(gnuplot) \space\space\space\@spaces}{%
      Package graphicx or graphics not loaded%
    }{See the gnuplot documentation for explanation.%
    }{The gnuplot epslatex terminal needs graphicx.sty or graphics.sty.}%
    \renewcommand\includegraphics[2][]{}%
  }%
  \providecommand\rotatebox[2]{#2}%
  \@ifundefined{ifGPcolor}{%
    \newif\ifGPcolor
    \GPcolorfalse
  }{}%
  \@ifundefined{ifGPblacktext}{%
    \newif\ifGPblacktext
    \GPblacktexttrue
  }{}%
  \let\gplgaddtomacro\g@addto@macro
  \gdef\gplbacktext{}%
  \gdef\gplfronttext{}%
  \makeatother
  \ifGPblacktext
    \def\colorrgb#1{}%
    \def\colorgray#1{}%
  \else
    \ifGPcolor
      \def\colorrgb#1{\color[rgb]{#1}}%
      \def\colorgray#1{\color[gray]{#1}}%
      \expandafter\def\csname LTw\endcsname{\color{white}}%
      \expandafter\def\csname LTb\endcsname{\color{black}}%
      \expandafter\def\csname LTa\endcsname{\color{black}}%
      \expandafter\def\csname LT0\endcsname{\color[rgb]{1,0,0}}%
      \expandafter\def\csname LT1\endcsname{\color[rgb]{0,1,0}}%
      \expandafter\def\csname LT2\endcsname{\color[rgb]{0,0,1}}%
      \expandafter\def\csname LT3\endcsname{\color[rgb]{1,0,1}}%
      \expandafter\def\csname LT4\endcsname{\color[rgb]{0,1,1}}%
      \expandafter\def\csname LT5\endcsname{\color[rgb]{1,1,0}}%
      \expandafter\def\csname LT6\endcsname{\color[rgb]{0,0,0}}%
      \expandafter\def\csname LT7\endcsname{\color[rgb]{1,0.3,0}}%
      \expandafter\def\csname LT8\endcsname{\color[rgb]{0.5,0.5,0.5}}%
    \else
      \def\colorrgb#1{\color{black}}%
      \def\colorgray#1{\color[gray]{#1}}%
      \expandafter\def\csname LTw\endcsname{\color{white}}%
      \expandafter\def\csname LTb\endcsname{\color{black}}%
      \expandafter\def\csname LTa\endcsname{\color{black}}%
      \expandafter\def\csname LT0\endcsname{\color{black}}%
      \expandafter\def\csname LT1\endcsname{\color{black}}%
      \expandafter\def\csname LT2\endcsname{\color{black}}%
      \expandafter\def\csname LT3\endcsname{\color{black}}%
      \expandafter\def\csname LT4\endcsname{\color{black}}%
      \expandafter\def\csname LT5\endcsname{\color{black}}%
      \expandafter\def\csname LT6\endcsname{\color{black}}%
      \expandafter\def\csname LT7\endcsname{\color{black}}%
      \expandafter\def\csname LT8\endcsname{\color{black}}%
    \fi
  \fi
    \setlength{\unitlength}{0.0500bp}%
    \ifx\gptboxheight\undefined%
      \newlength{\gptboxheight}%
      \newlength{\gptboxwidth}%
      \newsavebox{\gptboxtext}%
    \fi%
    \setlength{\fboxrule}{0.5pt}%
    \setlength{\fboxsep}{1pt}%
\begin{picture}(5102.00,6802.00)%
    \gplgaddtomacro\gplbacktext{%
      \csname LTb\endcsname%
      \put(1061,5305){\makebox(0,0)[r]{\strut{}$0$}}%
      \put(1061,5645){\makebox(0,0)[r]{\strut{}$0.2$}}%
      \put(1061,5985){\makebox(0,0)[r]{\strut{}$0.4$}}%
      \put(1061,6324){\makebox(0,0)[r]{\strut{}$0.6$}}%
      \put(1061,6664){\makebox(0,0)[r]{\strut{}$0.8$}}%
      \put(1193,5085){\makebox(0,0){\strut{}}}%
      \put(1872,5085){\makebox(0,0){\strut{}}}%
      \put(2552,5085){\makebox(0,0){\strut{}}}%
      \put(3231,5085){\makebox(0,0){\strut{}}}%
      \put(3911,5085){\makebox(0,0){\strut{}}}%
      \put(4590,5085){\makebox(0,0){\strut{}}}%
    }%
    \gplgaddtomacro\gplfronttext{%
      \csname LTb\endcsname%
      \put(423,5984){\rotatebox{-270}{\makebox(0,0){\strut{}$T(P)$ $(P_{\rm o}/R_\star)$}}}%
      \put(2891,5019){\makebox(0,0){\strut{}}}%
      \put(1659,6502){\makebox(0,0)[r]{\strut{}\footnotesize{$t=0$}}}%
      \csname LTb\endcsname%
      \put(1659,6348){\makebox(0,0)[r]{\strut{}\footnotesize{$500$}}}%
      \put(1659,6194){\makebox(0,0)[r]{\strut{}\footnotesize{$1000$}}}%
      \put(1659,6040){\makebox(0,0)[r]{\strut{}\footnotesize{$1500$}}}%
      \put(1659,5886){\makebox(0,0)[r]{\strut{}\footnotesize{$1579$}}}%
    }%
    \gplgaddtomacro\gplbacktext{%
      \csname LTb\endcsname%
      \put(1061,3741){\makebox(0,0)[r]{\strut{}$0$}}%
      \put(1061,4081){\makebox(0,0)[r]{\strut{}$0.5$}}%
      \put(1061,4421){\makebox(0,0)[r]{\strut{}$1$}}%
      \put(1061,4760){\makebox(0,0)[r]{\strut{}$1.5$}}%
      \put(1061,5100){\makebox(0,0)[r]{\strut{}$2$}}%
      \put(1193,3521){\makebox(0,0){\strut{}}}%
      \put(1872,3521){\makebox(0,0){\strut{}}}%
      \put(2552,3521){\makebox(0,0){\strut{}}}%
      \put(3231,3521){\makebox(0,0){\strut{}}}%
      \put(3911,3521){\makebox(0,0){\strut{}}}%
      \put(4590,3521){\makebox(0,0){\strut{}}}%
    }%
    \gplgaddtomacro\gplfronttext{%
      \csname LTb\endcsname%
      \put(423,4420){\rotatebox{-270}{\makebox(0,0){\strut{}$\varphi$ (rad)}}}%
      \put(2891,3455){\makebox(0,0){\strut{}}}%
    }%
    \gplgaddtomacro\gplbacktext{%
      \csname LTb\endcsname%
      \put(1059,3441){\makebox(0,0)[r]{\strut{}$1$}}%
      \put(1059,2176){\makebox(0,0)[r]{\strut{}$10^{-4}$}}%
      \put(1059,2492){\makebox(0,0)[r]{\strut{}$10^{-3}$}}%
      \put(1059,2808){\makebox(0,0)[r]{\strut{}$10^{-2}$}}%
      \put(1059,3125){\makebox(0,0)[r]{\strut{}$10^{-1}$}}%
      \put(1191,1956){\makebox(0,0){\strut{}}}%
      \put(1871,1956){\makebox(0,0){\strut{}}}%
      \put(2551,1956){\makebox(0,0){\strut{}}}%
      \put(3230,1956){\makebox(0,0){\strut{}}}%
      \put(3910,1956){\makebox(0,0){\strut{}}}%
      \put(4590,1956){\makebox(0,0){\strut{}}}%
    }%
    \gplgaddtomacro\gplfronttext{%
      \csname LTb\endcsname%
      \put(421,2856){\rotatebox{-270}{\makebox(0,0){\strut{}$v_{\varphi}$ (km/yr)}}}%
      \put(2890,1890){\makebox(0,0){\strut{}}}%
    }%
    \gplgaddtomacro\gplbacktext{%
      \csname LTb\endcsname%
      \put(1059,612){\makebox(0,0)[r]{\strut{}-0.6}}%
      \put(1059,952){\makebox(0,0)[r]{\strut{}-0.4}}%
      \put(1059,1292){\makebox(0,0)[r]{\strut{}-0.2}}%
      \put(1059,1632){\makebox(0,0)[r]{\strut{}0}}%
      \put(1059,1972){\makebox(0,0)[r]{\strut{}0.2}}%
      \put(1191,392){\makebox(0,0){\strut{}0.5}}%
      \put(1871,392){\makebox(0,0){\strut{}0.6}}%
      \put(2551,392){\makebox(0,0){\strut{}0.7}}%
      \put(3230,392){\makebox(0,0){\strut{}0.8}}%
      \put(3910,392){\makebox(0,0){\strut{}0.9}}%
      \put(4590,392){\makebox(0,0){\strut{}1}}%
    }%
    \gplgaddtomacro\gplfronttext{%
      \csname LTb\endcsname%
      \put(421,1292){\rotatebox{-270}{\makebox(0,0){\strut{}$V$ (MV)}}}%
      \put(2890,62){\makebox(0,0){\strut{}$P$ $(P_{\rm o})$}}%
    }%
    \gplbacktext
    \put(0,0){\includegraphics{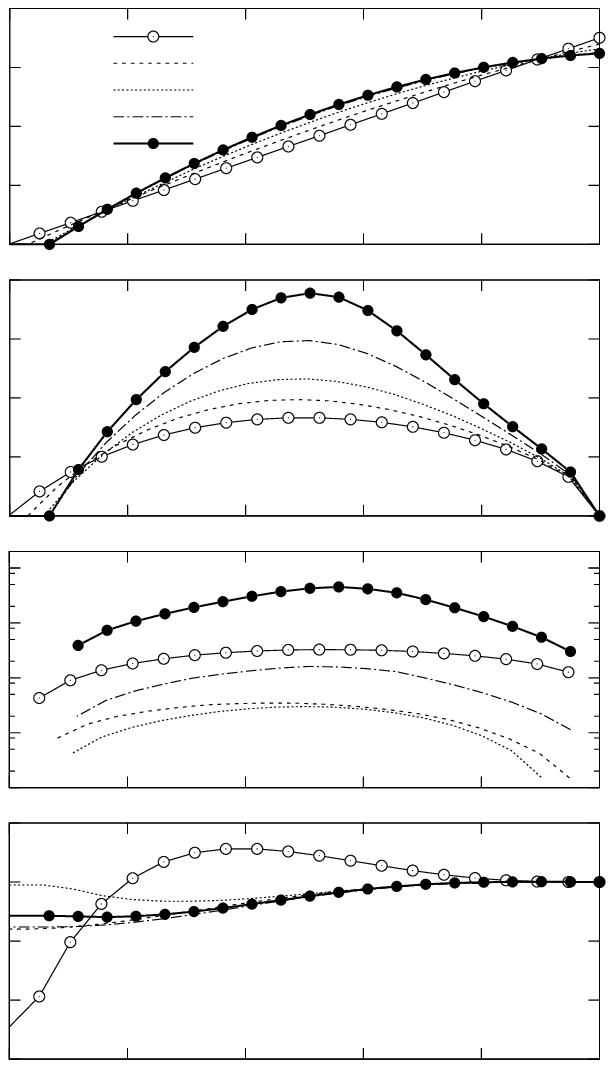}}%
    \gplfronttext
  \end{picture}%
\endgroup

%% file: figures/epslatex/pcritical.tex
\begingroup
  \makeatletter
  \providecommand\color[2][]{%
    \GenericError{(gnuplot) \space\space\space\@spaces}{%
      Package color not loaded in conjunction with
      terminal option `colourtext'%
    }{See the gnuplot documentation for explanation.%
    }{Either use 'blacktext' in gnuplot or load the package
      color.sty in LaTeX.}%
    \renewcommand\color[2][]{}%
  }%
  \providecommand\includegraphics[2][]{%
    \GenericError{(gnuplot) \space\space\space\@spaces}{%
      Package graphicx or graphics not loaded%
    }{See the gnuplot documentation for explanation.%
    }{The gnuplot epslatex terminal needs graphicx.sty or graphics.sty.}%
    \renewcommand\includegraphics[2][]{}%
  }%
  \providecommand\rotatebox[2]{#2}%
  \@ifundefined{ifGPcolor}{%
    \newif\ifGPcolor
    \GPcolorfalse
  }{}%
  \@ifundefined{ifGPblacktext}{%
    \newif\ifGPblacktext
    \GPblacktexttrue
  }{}%
  \let\gplgaddtomacro\g@addto@macro
  \gdef\gplbacktext{}%
  \gdef\gplfronttext{}%
  \makeatother
  \ifGPblacktext
    \def\colorrgb#1{}%
    \def\colorgray#1{}%
  \else
    \ifGPcolor
      \def\colorrgb#1{\color[rgb]{#1}}%
      \def\colorgray#1{\color[gray]{#1}}%
      \expandafter\def\csname LTw\endcsname{\color{white}}%
      \expandafter\def\csname LTb\endcsname{\color{black}}%
      \expandafter\def\csname LTa\endcsname{\color{black}}%
      \expandafter\def\csname LT0\endcsname{\color[rgb]{1,0,0}}%
      \expandafter\def\csname LT1\endcsname{\color[rgb]{0,1,0}}%
      \expandafter\def\csname LT2\endcsname{\color[rgb]{0,0,1}}%
      \expandafter\def\csname LT3\endcsname{\color[rgb]{1,0,1}}%
      \expandafter\def\csname LT4\endcsname{\color[rgb]{0,1,1}}%
      \expandafter\def\csname LT5\endcsname{\color[rgb]{1,1,0}}%
      \expandafter\def\csname LT6\endcsname{\color[rgb]{0,0,0}}%
      \expandafter\def\csname LT7\endcsname{\color[rgb]{1,0.3,0}}%
      \expandafter\def\csname LT8\endcsname{\color[rgb]{0.5,0.5,0.5}}%
    \else
      \def\colorrgb#1{\color{black}}%
      \def\colorgray#1{\color[gray]{#1}}%
      \expandafter\def\csname LTw\endcsname{\color{white}}%
      \expandafter\def\csname LTb\endcsname{\color{black}}%
      \expandafter\def\csname LTa\endcsname{\color{black}}%
      \expandafter\def\csname LT0\endcsname{\color{black}}%
      \expandafter\def\csname LT1\endcsname{\color{black}}%
      \expandafter\def\csname LT2\endcsname{\color{black}}%
      \expandafter\def\csname LT3\endcsname{\color{black}}%
      \expandafter\def\csname LT4\endcsname{\color{black}}%
      \expandafter\def\csname LT5\endcsname{\color{black}}%
      \expandafter\def\csname LT6\endcsname{\color{black}}%
      \expandafter\def\csname LT7\endcsname{\color{black}}%
      \expandafter\def\csname LT8\endcsname{\color{black}}%
    \fi
  \fi
    \setlength{\unitlength}{0.0500bp}%
    \ifx\gptboxheight\undefined%
      \newlength{\gptboxheight}%
      \newlength{\gptboxwidth}%
      \newsavebox{\gptboxtext}%
    \fi%
    \setlength{\fboxrule}{0.5pt}%
    \setlength{\fboxsep}{1pt}%
\begin{picture}(5102.00,2834.00)%
    \gplgaddtomacro\gplbacktext{%
      \csname LTb\endcsname%
      \put(814,725){\makebox(0,0)[r]{\strut{}$3.3$}}%
      \put(814,1333){\makebox(0,0)[r]{\strut{}$3.4$}}%
      \put(814,1940){\makebox(0,0)[r]{\strut{}$3.5$}}%
      \put(814,2548){\makebox(0,0)[r]{\strut{}$3.6$}}%
      \put(946,505){\makebox(0,0){\strut{}0}}%
      \put(1857,505){\makebox(0,0){\strut{}400}}%
      \put(2768,505){\makebox(0,0){\strut{}800}}%
      \put(3679,505){\makebox(0,0){\strut{}1200}}%
      \put(4590,505){\makebox(0,0){\strut{}1600}}%
    }%
    \gplgaddtomacro\gplfronttext{%
      \csname LTb\endcsname%
      \put(176,1636){\rotatebox{-270}{\makebox(0,0){\strut{}$P_{\rm c}$ \& $\dfrac{P_{\rm o}}{2}$ ($10^{25}$\,G\,cm$^2$)}}}%
      \put(2768,175){\makebox(0,0){\strut{}Time (yr)}}%
      \put(1458,2231){\makebox(0,0)[r]{\strut{}$P_{\rm c}$}}%
      \put(1458,1901){\makebox(0,0)[r]{\strut{}$\dfrac{P_{\rm o}}{2}$}}%
    }%
    \gplbacktext
    \put(0,0){\includegraphics{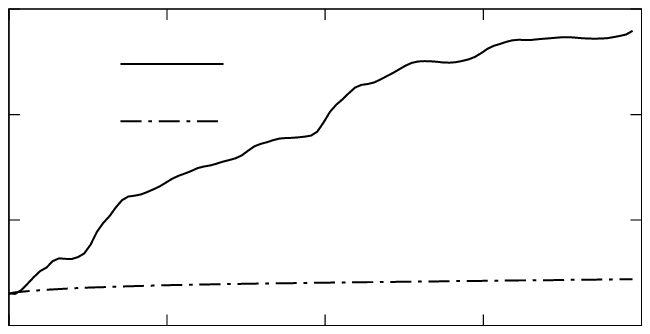}}%
    \gplfronttext
  \end{picture}%
\endgroup

%% file: figures/epslatex/t_profile.tex
\begingroup
  \makeatletter
  \providecommand\color[2][]{%
    \GenericError{(gnuplot) \space\space\space\@spaces}{%
      Package color not loaded in conjunction with
      terminal option `colourtext'%
    }{See the gnuplot documentation for explanation.%
    }{Either use 'blacktext' in gnuplot or load the package
      color.sty in LaTeX.}%
    \renewcommand\color[2][]{}%
  }%
  \providecommand\includegraphics[2][]{%
    \GenericError{(gnuplot) \space\space\space\@spaces}{%
      Package graphicx or graphics not loaded%
    }{See the gnuplot documentation for explanation.%
    }{The gnuplot epslatex terminal needs graphicx.sty or graphics.sty.}%
    \renewcommand\includegraphics[2][]{}%
  }%
  \providecommand\rotatebox[2]{#2}%
  \@ifundefined{ifGPcolor}{%
    \newif\ifGPcolor
    \GPcolorfalse
  }{}%
  \@ifundefined{ifGPblacktext}{%
    \newif\ifGPblacktext
    \GPblacktexttrue
  }{}%
  \let\gplgaddtomacro\g@addto@macro
  \gdef\gplbacktext{}%
  \gdef\gplfronttext{}%
  \makeatother
  \ifGPblacktext
    \def\colorrgb#1{}%
    \def\colorgray#1{}%
  \else
    \ifGPcolor
      \def\colorrgb#1{\color[rgb]{#1}}%
      \def\colorgray#1{\color[gray]{#1}}%
      \expandafter\def\csname LTw\endcsname{\color{white}}%
      \expandafter\def\csname LTb\endcsname{\color{black}}%
      \expandafter\def\csname LTa\endcsname{\color{black}}%
      \expandafter\def\csname LT0\endcsname{\color[rgb]{1,0,0}}%
      \expandafter\def\csname LT1\endcsname{\color[rgb]{0,1,0}}%
      \expandafter\def\csname LT2\endcsname{\color[rgb]{0,0,1}}%
      \expandafter\def\csname LT3\endcsname{\color[rgb]{1,0,1}}%
      \expandafter\def\csname LT4\endcsname{\color[rgb]{0,1,1}}%
      \expandafter\def\csname LT5\endcsname{\color[rgb]{1,1,0}}%
      \expandafter\def\csname LT6\endcsname{\color[rgb]{0,0,0}}%
      \expandafter\def\csname LT7\endcsname{\color[rgb]{1,0.3,0}}%
      \expandafter\def\csname LT8\endcsname{\color[rgb]{0.5,0.5,0.5}}%
    \else
      \def\colorrgb#1{\color{black}}%
      \def\colorgray#1{\color[gray]{#1}}%
      \expandafter\def\csname LTw\endcsname{\color{white}}%
      \expandafter\def\csname LTb\endcsname{\color{black}}%
      \expandafter\def\csname LTa\endcsname{\color{black}}%
      \expandafter\def\csname LT0\endcsname{\color{black}}%
      \expandafter\def\csname LT1\endcsname{\color{black}}%
      \expandafter\def\csname LT2\endcsname{\color{black}}%
      \expandafter\def\csname LT3\endcsname{\color{black}}%
      \expandafter\def\csname LT4\endcsname{\color{black}}%
      \expandafter\def\csname LT5\endcsname{\color{black}}%
      \expandafter\def\csname LT6\endcsname{\color{black}}%
      \expandafter\def\csname LT7\endcsname{\color{black}}%
      \expandafter\def\csname LT8\endcsname{\color{black}}%
    \fi
  \fi
    \setlength{\unitlength}{0.0500bp}%
    \ifx\gptboxheight\undefined%
      \newlength{\gptboxheight}%
      \newlength{\gptboxwidth}%
      \newsavebox{\gptboxtext}%
    \fi%
    \setlength{\fboxrule}{0.5pt}%
    \setlength{\fboxsep}{1pt}%
\begin{picture}(5102.00,2834.00)%
    \gplgaddtomacro\gplbacktext{%
      \csname LTb\endcsname%
      \put(828,704){\makebox(0,0)[r]{\strut{}$0$}}%
      \put(828,1170){\makebox(0,0)[r]{\strut{}$0.2$}}%
      \put(828,1637){\makebox(0,0)[r]{\strut{}$0.4$}}%
      \put(828,2103){\makebox(0,0)[r]{\strut{}$0.6$}}%
      \put(828,2569){\makebox(0,0)[r]{\strut{}$0.8$}}%
      \put(960,484){\makebox(0,0){\strut{}$0$}}%
      \put(1706,484){\makebox(0,0){\strut{}$0.2$}}%
      \put(2452,484){\makebox(0,0){\strut{}$0.4$}}%
      \put(3199,484){\makebox(0,0){\strut{}$0.6$}}%
      \put(3945,484){\makebox(0,0){\strut{}$0.8$}}%
      \put(4691,484){\makebox(0,0){\strut{}$1$}}%
    }%
    \gplgaddtomacro\gplfronttext{%
      \csname LTb\endcsname%
      \put(190,1636){\rotatebox{-270}{\makebox(0,0){\strut{}$T(P)$ $(P_{\rm o}/R_\star)$}}}%
      \put(2825,154){\makebox(0,0){\strut{}$(P - P_{\rm c})/(P_{\rm o} - P_{\rm c})$}}%
      \put(1572,2369){\makebox(0,0)[r]{\strut{}$\mathfrak{f} = 0.6$}}%
      \put(1572,2204){\makebox(0,0)[r]{\strut{}$0.5$}}%
      \put(1572,2039){\makebox(0,0)[r]{\strut{}$0.4$}}%
      \put(1572,1874){\makebox(0,0)[r]{\strut{}$0.3$}}%
      \put(1572,1709){\makebox(0,0)[r]{\strut{}$0.2$}}%
      \put(1572,1544){\makebox(0,0)[r]{\strut{}$0.1$}}%
    }%
    \gplbacktext
    \put(0,0){\includegraphics{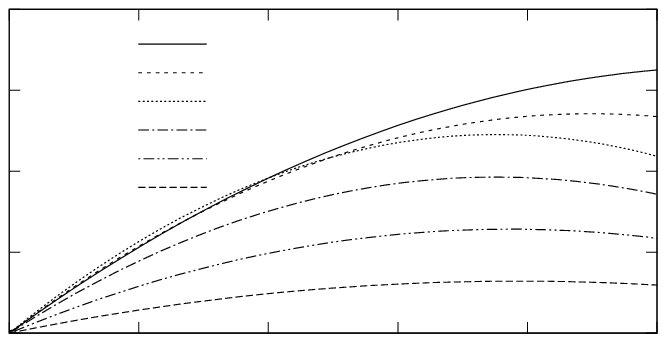}}%
    \gplfronttext
  \end{picture}%
\endgroup

%% file: figures/epslatex/spindown.tex
\begingroup
  \makeatletter
  \providecommand\color[2][]{%
    \GenericError{(gnuplot) \space\space\space\@spaces}{%
      Package color not loaded in conjunction with
      terminal option `colourtext'%
    }{See the gnuplot documentation for explanation.%
    }{Either use 'blacktext' in gnuplot or load the package
      color.sty in LaTeX.}%
    \renewcommand\color[2][]{}%
  }%
  \providecommand\includegraphics[2][]{%
    \GenericError{(gnuplot) \space\space\space\@spaces}{%
      Package graphicx or graphics not loaded%
    }{See the gnuplot documentation for explanation.%
    }{The gnuplot epslatex terminal needs graphicx.sty or graphics.sty.}%
    \renewcommand\includegraphics[2][]{}%
  }%
  \providecommand\rotatebox[2]{#2}%
  \@ifundefined{ifGPcolor}{%
    \newif\ifGPcolor
    \GPcolorfalse
  }{}%
  \@ifundefined{ifGPblacktext}{%
    \newif\ifGPblacktext
    \GPblacktexttrue
  }{}%
  \let\gplgaddtomacro\g@addto@macro
  \gdef\gplbacktext{}%
  \gdef\gplfronttext{}%
  \makeatother
  \ifGPblacktext
    \def\colorrgb#1{}%
    \def\colorgray#1{}%
  \else
    \ifGPcolor
      \def\colorrgb#1{\color[rgb]{#1}}%
      \def\colorgray#1{\color[gray]{#1}}%
      \expandafter\def\csname LTw\endcsname{\color{white}}%
      \expandafter\def\csname LTb\endcsname{\color{black}}%
      \expandafter\def\csname LTa\endcsname{\color{black}}%
      \expandafter\def\csname LT0\endcsname{\color[rgb]{1,0,0}}%
      \expandafter\def\csname LT1\endcsname{\color[rgb]{0,1,0}}%
      \expandafter\def\csname LT2\endcsname{\color[rgb]{0,0,1}}%
      \expandafter\def\csname LT3\endcsname{\color[rgb]{1,0,1}}%
      \expandafter\def\csname LT4\endcsname{\color[rgb]{0,1,1}}%
      \expandafter\def\csname LT5\endcsname{\color[rgb]{1,1,0}}%
      \expandafter\def\csname LT6\endcsname{\color[rgb]{0,0,0}}%
      \expandafter\def\csname LT7\endcsname{\color[rgb]{1,0.3,0}}%
      \expandafter\def\csname LT8\endcsname{\color[rgb]{0.5,0.5,0.5}}%
    \else
      \def\colorrgb#1{\color{black}}%
      \def\colorgray#1{\color[gray]{#1}}%
      \expandafter\def\csname LTw\endcsname{\color{white}}%
      \expandafter\def\csname LTb\endcsname{\color{black}}%
      \expandafter\def\csname LTa\endcsname{\color{black}}%
      \expandafter\def\csname LT0\endcsname{\color{black}}%
      \expandafter\def\csname LT1\endcsname{\color{black}}%
      \expandafter\def\csname LT2\endcsname{\color{black}}%
      \expandafter\def\csname LT3\endcsname{\color{black}}%
      \expandafter\def\csname LT4\endcsname{\color{black}}%
      \expandafter\def\csname LT5\endcsname{\color{black}}%
      \expandafter\def\csname LT6\endcsname{\color{black}}%
      \expandafter\def\csname LT7\endcsname{\color{black}}%
      \expandafter\def\csname LT8\endcsname{\color{black}}%
    \fi
  \fi
    \setlength{\unitlength}{0.0500bp}%
    \ifx\gptboxheight\undefined%
      \newlength{\gptboxheight}%
      \newlength{\gptboxwidth}%
      \newsavebox{\gptboxtext}%
    \fi%
    \setlength{\fboxrule}{0.5pt}%
    \setlength{\fboxsep}{1pt}%
\begin{picture}(5102.00,3400.00)%
    \gplgaddtomacro\gplbacktext{%
      \csname LTb\endcsname%
      \put(1059,1971){\makebox(0,0)[r]{\strut{}$0.8$}}%
      \put(1059,2311){\makebox(0,0)[r]{\strut{}$1$}}%
      \put(1059,2651){\makebox(0,0)[r]{\strut{}$1.2$}}%
      \put(1059,2991){\makebox(0,0)[r]{\strut{}$1.4$}}%
      \put(1059,3331){\makebox(0,0)[r]{\strut{}$1.6$}}%
      \put(1191,1751){\makebox(0,0){\strut{}}}%
      \put(2041,1751){\makebox(0,0){\strut{}}}%
      \put(2891,1751){\makebox(0,0){\strut{}}}%
      \put(3740,1751){\makebox(0,0){\strut{}}}%
      \put(4590,1751){\makebox(0,0){\strut{}}}%
    }%
    \gplgaddtomacro\gplfronttext{%
      \csname LTb\endcsname%
      \put(421,2651){\rotatebox{-270}{\makebox(0,0){\strut{}$\cal{P \dot{P}}$ ($10^{-11}$\,s)}}}%
      \put(2890,1685){\makebox(0,0){\strut{}}}%
    }%
    \gplgaddtomacro\gplbacktext{%
      \csname LTb\endcsname%
      \put(1061,408){\makebox(0,0)[r]{\strut{}-4}}%
      \put(1061,748){\makebox(0,0)[r]{\strut{}-2}}%
      \put(1061,1088){\makebox(0,0)[r]{\strut{}0}}%
      \put(1061,1427){\makebox(0,0)[r]{\strut{}2}}%
      \put(1061,1767){\makebox(0,0)[r]{\strut{}4}}%
      \put(1193,188){\makebox(0,0){\strut{}0}}%
      \put(2042,188){\makebox(0,0){\strut{}400}}%
      \put(2892,188){\makebox(0,0){\strut{}800}}%
      \put(3741,188){\makebox(0,0){\strut{}1200}}%
      \put(4590,188){\makebox(0,0){\strut{}1600}}%
    }%
    \gplgaddtomacro\gplfronttext{%
      \csname LTb\endcsname%
      \put(423,1087){\rotatebox{-270}{\makebox(0,0){\strut{}$n$}}}%
      \put(2891,-142){\makebox(0,0){\strut{}Time (yr)}}%
    }%
    \gplbacktext
    \put(0,0){\includegraphics{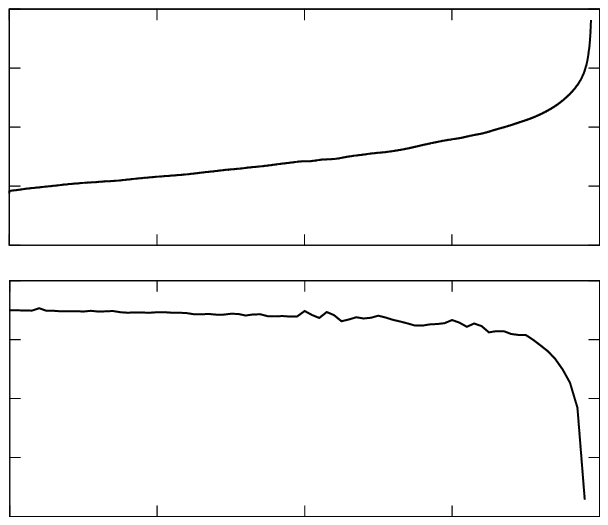}}%
    \gplfronttext
  \end{picture}%
\endgroup